\documentclass[twocolumn]{aastex701}
\usepackage{hyperref}
\usepackage{comment}
\usepackage{rotating}

\newcommand{\largefig}[1]{#1}

\begin{document}


\title{Dust-Embedded Star Formation:\\ Bridging Magellanic Cloud Studies of Massive Young Stellar Objects to Nearby Spiral Galaxies}

\author[0000-0002-0579-6613]{M. Jimena Rodríguez}
\affiliation{Space Telescope Science Institute, 3700 San Martin Drive, Baltimore, MD 21218, USA}
\affiliation{Instituto de Astrofísica de La Plata, CONICET--UNLP, Paseo del Bosque S/N, B1900FWA La Plata, Argentina }
\email{jrodriguez@stsci.edu}

\author[0000-0002-4663-6827]{R\'emy Indebetouw}
\affiliation{University of Virginia Astronomy Department, P.O. Box 400325, Charlottesville, VA, 22904, USA}
\affiliation{National Radio Astronomy Observatory, 520 Edgemont Rd, Charlottesville, VA 22903, USA}
\email{remy@virginia.edu}

\author[0000-0002-2278-9407]{Janice C. Lee}
\affiliation{Space Telescope Science Institute, 3700 San Martin Drive, Baltimore, MD 21218, USA}
\email{jlee@stsci.edu}

\author[0000-0002-3784-7032]{Bradley C. Whitmore}
\affiliation{Space Telescope Science Institute, 3700 San Martin Drive, Baltimore, MD 21218, USA}
\email{whitmore@stsci.edu}

\author[0000-0002-8528-7340]{David A. Thilker}
\affiliation{Department of Physics and Astronomy, The Johns Hopkins University, Baltimore, MD 21218, USA}
\email{dthilker@pha.jhu.edu}

\author[0000-0002-5937-9778]{J. Peltonen}
\affiliation{Department of Physics, University of Alberta, Edmonton, AB T6G 2E1, Canada}
\email{peltonen@ualberta.ca}

\author[0000-0002-5204-2259]{Erik Rosolowsky}
\affiliation{Department of Physics, University of Alberta, Edmonton, AB T6G 2E1, Canada}
\email{rosolowsky@ualberta.ca}

\author[0000-0002-0012-2142]{Thomas G. Williams}
\affiliation{UK ALMA Regional Centre Node, Jodrell Bank Centre for Astrophysics, Department of Physics and Astronomy, The University of Manchester, Oxford Road, Manchester M13 9PL, UK}
\email{thomas.g.williams94@gmail.com}

\author[0000-0001-9605-780X]{Eric W. Koch}
\affiliation{National Radio Astronomy Observatory, 520 Edgemont Rd, Charlottesville, VA 22903, USA}
\email{koch.eric.w@gmail.com}

\author[0000-0002-1723-6330]{Bruce Elmegreen}
\email{belmegreen@gmail.com}
\affiliation{Katonah, NY 10536}

\author[0000-0002-0560-3172]{Ralf S.\ Klessen}
\affiliation{Universit\"{a}t Heidelberg, Zentrum f\"{u}r Astronomie, Institut f\"{u}r Theoretische Astrophysik, Albert-Ueberle-Str.\ 2, 69120 Heidelberg, Germany}
\affiliation{Universit\"{a}t Heidelberg, Interdisziplin\"{a}res Zentrum f\"{u}r Wissenschaftliches Rechnen, Im Neuenheimer Feld 225, 69120 Heidelberg, Germany}
\email{klessen@structures.uni-heidelberg.de}

\author[0000-0002-5158-243X]{Roberta Paladini}
\affiliation{IPAC, Caltech, Pasadena, California 91125, USA}
\email{paladini@ipac.caltech.edu}

\author[0000-0002-4781-7291]{Sumit Sarbadhicary}
\affiliation{Department of Physics and Astronomy, The Johns Hopkins University, Baltimore, MD 21218, USA}
\email{ssarbad1@jh.edu}

\author[0000-0002-5782-9093]{Daniel~A.~Dale}
\affiliation{Department of Physics and Astronomy, University of Wyoming, Laramie, WY 82071, USA}
\email{ddale@uwyo.edu}

\author[0000-0001-8348-2671]{Kelsey E. Johnson}
\affiliation{University of Virginia Astronomy Department, P.O. Box 400325, Charlottesville, VA, 22904, USA}
\email{kej7a@virginia.edu}

\author[0000-0001-8289-3428]{Aida Wofford}
\affiliation{Instituto de Astronom\'ia, Universidad Nacional Aut\'onoma de M\'exico, Unidad Acad\'emica en Ensenada, Km 103 Carr. Tijuana$-$Ensenada, Ensenada, B.C.,\\C.P. 22860, M\'exico}
\email{awofford@astro.unam.mx}



\begin{abstract}
We use JWST NIRCam and MIRI imaging at 2, 4, 10, and 21~$\mu$m to study young, dusty compact sources in four nearby galaxies at distances of $\sim$1–5~Mpc (M33, NGC~300, NGC~7793, and NGC~5068). This work bridges well-characterized massive young stellar objects (MYSOs) in the Magellanic Clouds from the \textit{Spitzer} SAGE survey to new studies of embedded clusters in more distant galaxies with JWST. Guided by the SAGE-LMC catalog, we define JWST color–magnitude selection criteria  ($\rm F1000W$ versus $\rm F1000W-F2100W$) and test them using resolution-degradation experiments.
We identify 216, 32, 80 and 139
dusty young objects in the four galaxies, respectively.
The selected population spans sources from systems dominated by a single MYSO to compact marginally resolved sources hosting multiple MYSOs.
The color selection remains stable across 1–5~Mpc, and the 10$\mu$m luminosity function retains a slope of $\alpha\sim-2$. However, blending and surface-brightness dilution remove fainter sources, leading to incompleteness of up to $\sim$50\% at 5.2~Mpc and biasing the sample toward brighter objects (F1000W~$<$~19~mag). The sample spans approximate stellar masses of $\sim$10–$\rm 2\times 10^{5}~M_\odot$.
Spatial resolution affects the interpretation of mid-infrared emission: clustering increases the fraction of emission attributed to compact sources in active regions, while blending into diffuse emission dominates in quiescent environments. 
Comparisons with PAH-selected young clusters in the PHANGS galaxy NGC~5068 show that our selection recovers $\sim$80\% of the PAH-selected sources.
We show that the practical limit for studying individual MYSOs with JWST is $\sim$3~Mpc.  The resulting catalog provides a foundation for future resolved studies of star formation rates and early cluster evolution.

\end{abstract}

\keywords{\uat{Young stellar objects}{1834} --- \uat{Young star clusters}{1833} --- \uat{Star formation}{1569} --- \uat{Infrared astronomy}{786} --- \uat{James Webb Space Telescope}{2291}}


\section{Introduction} 
\label{sec:Intro}

Resolved studies of dust-obscured massive star formation provide essential ground truth for understanding the processes that drive star formation and galaxy evolution. Massive stars, although representing only a small fraction of stars formed, dominate the production of heavy elements and the injection of energy and momentum into the interstellar medium (ISM), thereby shaping the evolution of their host galaxies. Much of this influence occurs during a short phase lasting only $\sim$2--3~Myr, when massive stars complete most of their mass accretion and interact most strongly with their natal material through radiation pressure, stellar winds, and ionizing radiation. The study of massive young stellar objects (MYSOs) provides the most detailed view of the earliest stage of massive star formation, before these systems emerge as main-sequence O and B stars.

Where spatial resolution in the infrared permits, the goal is to obtain a complete census of individual YSOs and embedded clusters, with constraints on their ages and masses \citep[e.g.,][]{ladalada03, whitney08, gruendl09, Rodriguez23}. Not only are such studies needed to understand the earliest phases of star and cluster evolution; they provide an empirical foundation for calibrating integrated star-formation diagnostics used in more distant galaxies \citep[e.g.,][]{Heiderman10, Robitaille10, knutas25}.  In practice, however, these measurements are challenging: stellar age and mass estimates become highly uncertain when ultraviolet and optical emission from stellar photospheres is heavily extinguished, and separating compact sources from diffuse ISM emission becomes increasingly difficult as spatial resolution degrades at longer infrared wavelengths and larger distances. Recent studies have also highlighted the challenges associated with modeling the NIR emission of young stellar systems and deriving robust physical properties in nearby galaxies \citep{Henny25, Pedrini25}.

Prior to JWST, detailed studies of individual MYSOs were largely limited to the Milky Way and the Magellanic Clouds. The \textit{Spitzer} SAGE \citep{sage} and HERITAGE \citep{heritage} surveys enabled the first comprehensive census of dusty MYSOs with masses $\gtrsim 5~M_\odot$ \citep{whitney08,gruendl09,seale14}. These masses were estimated by fitting radiative transfer models of single massive stars with dusty envelopes \citep{Robitaille2006}, with mass primarily constrained by the bolometric luminosity. Far-infrared measurements extending to $\lambda \sim 100~\mu$m were essential for locating the peak of the spectral energy distribution; when sources could be measured only at shorter wavelengths ($\lambda \lesssim 20~\mu$m) not reaching the SED peak, the resulting mass estimates were substantially more uncertain \citep[e.g.][]{oneill22}. Although these Magellanic MYSOs likely contain more than one star, the steep (L$\sim$M$^{3-3.5}$) mass-luminosity relation for single stars ensures that the emission is dominated by the most massive star in each small group.

The small physical scale of embedded star formation further limited such work beyond the Local Group. Massive stars typically form in compact clusters with effective radii of $\sim$0.5--10~pc \citep[e.g.,][]{2010ARA&A..48..431P,2017ApJ...841...92R}. While the resolution of \textit{Spitzer} was sufficient to resolve such scales in the Magellanic Clouds, it corresponded to tens of parsecs in galaxies at distances of $\sim$1~Mpc, blending individual star-forming regions into unresolved complexes.

 With its transformative sensitivity and resolution, JWST now extends resolved studies of compact, dusty, and often optically faint sources to galaxies out to tens of megaparsecs across a wide range of environments \citep[e.g.,][]{Rodriguez23,whitmore23,linden24,pedrini24,Peltonen2024,lenkic24,Hassani25,Rodriguez25}. At distances of $\sim$1 Mpc, JWST achieves physical resolutions of $\sim$0.7--3 pc at near- to mid-infrared wavelengths, comparable to the scales which enabled the \textit{Spitzer} studies of MYSOs in the Magellanic Clouds. This capability thus potentially allows MYSO populations to be studied at parsec scales in galaxies beyond the Local Group, across a greater diversity of galactic environments.

However, methods for identifying and characterizing MYSOs with JWST have not yet been systematically validated across different distances and spatial resolutions, leaving uncertainties in classification reliability and completeness as a function of scale. In this paper we take initial steps toward establishing such methods, using the \textit{Spitzer} SAGE-LMC survey as a guide. We analyze JWST NIRCam and MIRI imaging at 2, 4.4, 10, and 21~$\mu$m to identify and classify young, dusty compact sources in four nearby spiral galaxies spanning distances of $\sim$1–5~Mpc (M33, NGC~300, NGC~7793, and NGC~5068). We examine how mid-infrared source selection changes with spatial resolution and assess the distances to which individual MYSOs can be studied with JWST. These results provide a first framework for interpreting JWST surveys of embedded star formation along a continuum of distances beyond the Local Group, and provide a candidate sample of young dusty compact sources (with sizes $\lesssim 2$ times the JWST resolution) in spiral galaxies at distances of $\sim$1--5~Mpc for future analysis.

The remainder of this paper is organized as follows. Section~2 describes the JWST observations used in this work, together with \textit{HST}, \textit{Spitzer}, and other ancillary datasets. In Section~3 we summarize the procedures for source detection, photometry, and the convolution of the JWST images to approximate observations at larger distances. The methods used for the selection and classification of young compact dusty sources, guided by the \textit{Spitzer} SAGE YSO dataset, are presented in Section~4. Section~5 examines how increasing distance and decreasing spatial resolution affect both the recovery of compact dusty sources and the fraction of mid-infrared emission attributed to compact star-forming regions versus diffuse dust heating. In Section~6 we present the luminosity function of the selected dusty population and obtain approximate mass estimates for the sources. Section~7 compares our catalogs with previous studies, including literature catalogs of dusty young objects in M~33 and recent PHANGS analyses of embedded clusters in NGC~5068. Section~8 summarizes our main results and their implications for identifying embedded star formation in nearby galaxies using mid-infrared observations.

\section{Data} 

This analysis is primarily based on observations from JWST Cycle 1 program GO-2130 (PI J. C. Lee), which was allocated $\sim$30 hours to study MYSOs in spiral arm features across 3 nearby galaxies (M33, NGC~300, NGC~7793, Fig.~\ref{fig:footprints}). The target galaxies were chosen to be at $\sim$1, 2, and 3~Mpc to allow us to understand limitations associated with MYSO studies due to increased blending with distance.  To this sample, we add a fourth spiral galaxy, NGC~5068, the nearest target in the PHANGS-JWST Cycle 1 Treasury Program \citep[GO-2107, PI J. C. Lee; see][]{phangs-jwst} to extend our blending analysis to $\sim$5~Mpc and examine how MYSO dominated populations overlap with the inventories of dust embedded clusters being conducted in more distant Local Volume galaxies \citep[e.g.,][]{Rodriguez25}.

\subsection{JWST Imaging Observations for M33, NGC 300, NGC 7793}

{\it Filters:} Building on methods established by Spitzer surveys of the Magellanic Clouds as discussed in the introduction, we identify MYSOs using NIRCam and MIRI four-band photometry at 2, 4.4, 10, and 21$\mu$m, which can distinguish MYSO-dominated clusters from other IR-bright sources. The 21$\mu$m flux traces dust continuum, while the flux at 10$\mu$m also samples continuum, but can potentially detect the presence of silicates which evolves from absorption to emission with MYSO stage \citep{seale09, Jones2017}. The 4.4~$\mu$m flux provides another point on the dust continuum, and 2~$\mu$m imaging yields the least obscured view of the stellar photospheric continuum at exquisite resolution (PSF FWHM 0\farcs066; Table~\ref{Tab:bands}).  The PHANGS-JWST Cycle~1 imaging of NGC~5068 obtained comparable imaging at 2, 3.6, 10, and 21$\mu$m, along with additional filters, as described further below.

\begin{figure*}
    \centering
    \includegraphics[width=1\linewidth]{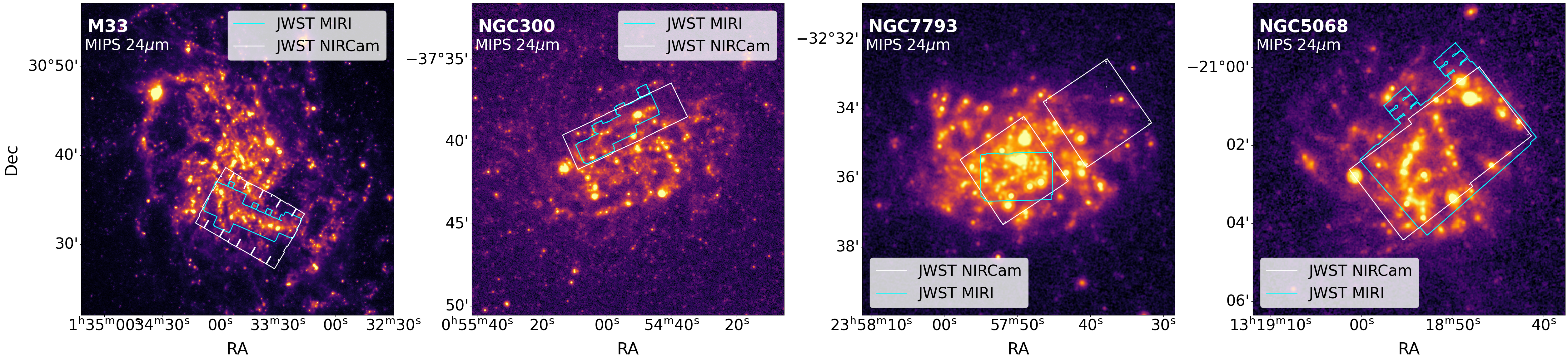}
    \caption{JWST NIRCam and MIRI observed footprints overlaid on a Spitzer MIPS 24$\mu$m image of each galaxy. M33, NGC~300, NGC~7793 were observed as part of GO-2130 (PI J.C. Lee), and NGC~5068 as part of the Cycle 1 PHANGS-JWST Treasury Survey \citep[GO-2107][]{phangs-jwst}.}
    \label{fig:footprints}
\end{figure*}

{\it Survey Footprints:} The size of the survey areas is $\sim$2--3~$\mathrm{kpc}^2$
(Fig.~\ref{fig:footprints}) for M33, NGC~300, NGC~7793. Areas along spiral arms were chosen to yield a statistically meaningful sample of embedded sources in each galaxy, and enable studies of their spatial distributions and timescales. The regions were also selected to overlap with archival HST and ALMA data, and allow comparison with molecular clouds and optically bright young clusters.
We obtained 15, 3, and 1 MIRI pointing(s) in M~33, NGC~300, and NGC~7793, respectively.  With MIRI, the F1000W and F2100W filters were used for all targets. With NIRCam, we obtained imaging in F200W and F444W for NGC~300 and NGC~7793. For M33, we instead used the medium-band filters F210M and F430M to mitigate against potential saturation from bright objects.

We use a MIRI 4-point dither pattern optimized for analysis of point sources.\footnote{\url{https://jwst-docs.stsci.edu/jwst-mid-infrared-instrument/miri-observing-strategies/miri-imaging-recommended-strategies\#MIRIImagingRecommendedStrategies-choosingdithersChoosingaditherpattern}} For NIRCAM, the INTRAMODULEBOX4 primary dither is used to maximize the area observed at full depth in NGC~300 and NGC~7793.  Since the target region is much larger in M~33, we use 
FULL/3TIGHT
for the primary dither pattern, with 2 subpixel positions to build depth.\footnote{\url{https://jwst-docs.stsci.edu/jwst-near-infrared-camera/nircam-observing-strategies/nircam-imaging-recommended-strategies\#gsc.tab=0}} In general, subpixel dithering is unnecessary because the PSF is well-sampled at F200W and F444W.  

Mosaicking in M33 employs a MIRI 6x4 grid, and within this grid 15 tiles capture star formation along the arm. For NGC~300 a 3x1 MIRI mosaic is used to cover the spiral arm. NGC~7793 is covered in a single MIRI pointing. For MIRI mosaics, we use 10\% tile overlap, and a 14\% row shift (in the case of NGC~300). NIRCAM mosaicking is needed in M33 (5x1 grid) and NGC~300 (1x2 grid). In M33, row overlap is 8\%; in NGC~300, column overlap is 58\%. The latter provides contiguous coverage in NGC~300 along the mosaic column-oriented extent.  Orient constraints were imposed to keep the JWST footprint on the area of interest.

\textit{Exposure times and read-out:} 
The total integration time for each galaxy and band are listed in Table~\ref{Tab:bands}.
To enable detection of faint sources without exceeding bright limits, we adopt a hybrid approach for both NIRCAM and MIRI imaging. For
NIRCAM we combine short exposures (with the fastest readout, RAPID) and longer observations (with slower readouts, BRIGHT2) to maximize dynamic range.  For MIRI we use FULL/FAST/Ngroups=5, but also target two bright areas in M~33 with short/fast readout subarray (SUB256) observations. 

The JWST images were processed using \texttt{pjpipe} \citep{2024Williams}, a reduction pipeline designed for PHANGS–JWST data. This software provides standardized calibration, mosaicking, and artifact mitigation optimized for extragalactic studies.

\subsection{Archival HST Data for M~33, NGC~300, NGC~7793}

We also used archival HST data. For M~33 we used data from the Panchromatic Hubble Andromeda Treasury: Triangulum Extended Region ``PHATTER",\footnote{https://archive.stsci.edu/hlsp/phatter}), which cover the near ultraviolet to the near-IR: F275W, F336W, F475W, F814W, F160W \citep{Phatter, https://doi.org/10.17909/t9-ksyp-na40}. The optical filters (F475W, F814W) were observed with the Advanced Camera for Surveys (ACS), while F275W, F336W and F160W were taken with Wide Field Camera 3 (WFC3), see \cite{Phatter} for a detailed description of the observing strategy.  

For NGC~300, we downloaded from the Barbara A. Mikulski Archive for Space Telescopes (MAST\footnote{\url{https://mast.stsci.edu/}}) all available HST observations overlapping with our JWST field of view (see Fig.\ref{fig:footprints}). These include data from several different programs (\#9492,\#10915,\#12450 and \#13515). We obtained a reasonable level of coverage across the field in the following bands: B (F435W/F475W), V (F555W/F606W), and I (F814W).

For NGC~7793 we used data from the Legacy ExtraGalactic UV Survey\footnote{https://archive.stsci.edu/prepds/legus/}) \citep[LEGUS][]{Legus,https://doi.org/10.17909/t9j01z}, which covers 5 broad bands from the NUV to the I (F275W, F336W, F438W, F555W and F814W).

\begin{figure*}
    \centering
    \includegraphics[width=1\linewidth]{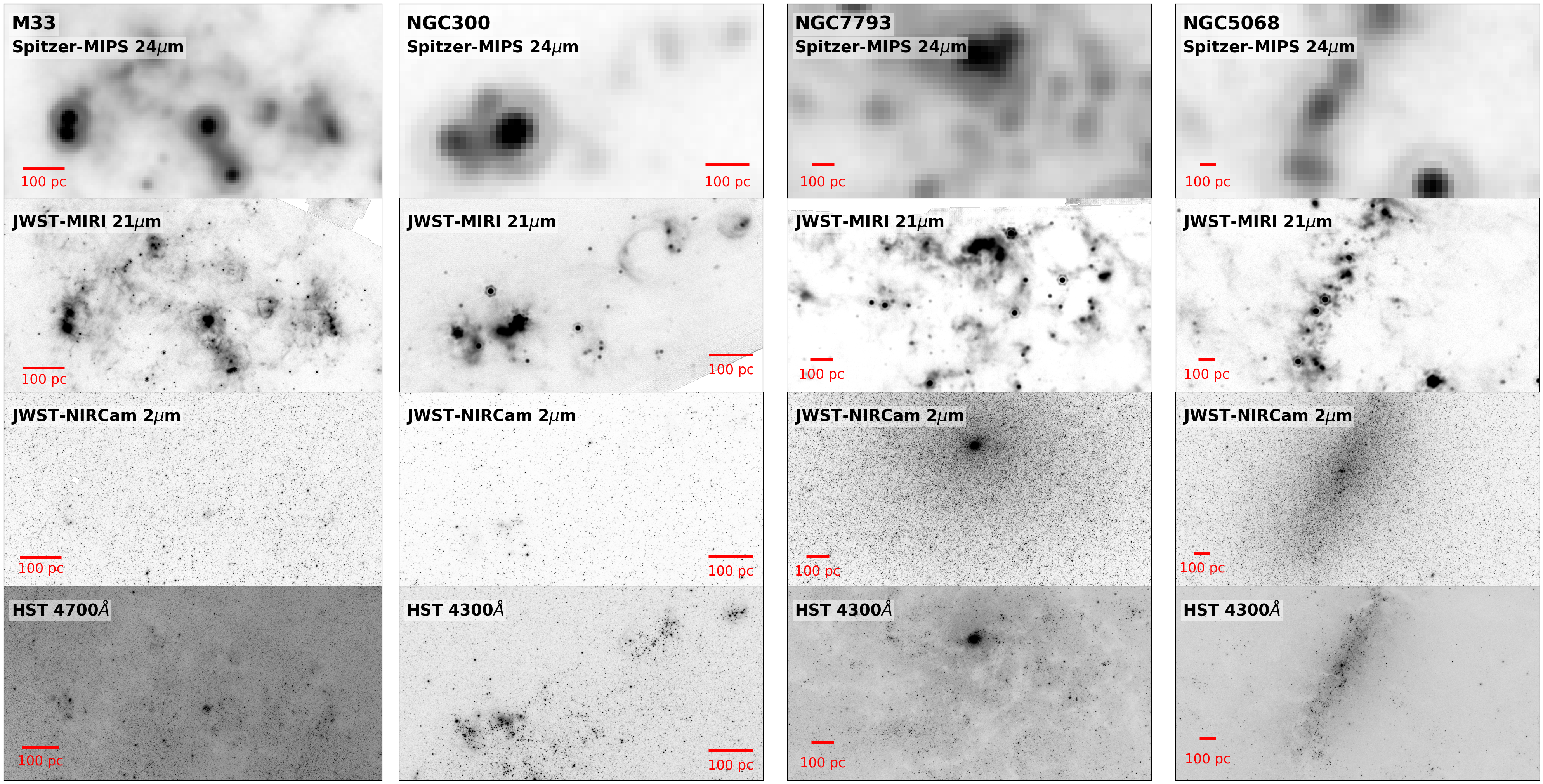}
    \caption{Comparison of JWST, HST, and Spitzer imaging for select areas of M~33, NGC~300, NGC~7793 and NGC~5068 to a provide a more detailed view of several star-forming regions in each galaxy.  Spitzer MIPS 24$\mu$m imaging is shown in the top panels. NIRCam 2$\mu$m and MIRI 21$\mu$m imaging shown in second and third row of panels respectively. The dramatic improvement in resolution and sensitivity achieved by JWST, which enables studies of massive YSOs and embedded clusters in more distant galaxies for the first time, is clear. The bottom row of panels shows the HST B-band image of the same area in each galaxy, and provides a view of the emission from young stars as well as obscuration by dust lanes and patches. 
    }
    \label{fig:24-B}
\end{figure*}

\subsection{NGC~5068: Archival JWST and HST data}

To our sample of galaxies, we added NGC~5068, the nearest galaxy in the PHANGS--JWST survey \citep[program 2107][]{phangs-jwst}, located at a distance of 5.2~Mpc. The JWST observations include eight filters: four NIRCam bands (F200W, F300M, F335M, and F360M) and four MIRI bands 
\citep[F770W, F1000W, F1130W, and F2100W,][]{https://doi.org/10.17909/ew88-jt15}. 
In this case, we benefit from a larger observed area (albeit at shallower depth), as the larger distances of the PHANGS sample allow the observations to cover most of the main body of each galaxy rather than only portions of its spiral arms.
NGC~5068 is also part of the PHANGS--HST survey \citep{phangs_HST}, with available HST data in five broad-band filters \cite[F275W, F336W, F438W, F555W, and F814W, ][]{https://doi.org/10.17909/t9-r08f-dq31}.

\subsection{Spitzer SAGE Survey and MYSO Catalogs}
We lean heavily on the \textit{Spitzer} SAGE survey of the Large Magellanic Cloud (LMC) to develop an empirical guide for identifying dusty young stellar populations with JWST observations. The SAGE and HERITAGE infrared surveys \citep{sage, https://doi.org/10.26131/irsa404,heritage}, combined with 2MASS \citep{2MASS} imaged the LMC 
at 1.1, 1.6, 2.2, 3.6, 4.5, 5.8, 8.0, 24, 70, 100, 160, 250, 350, and 500 $\mu$m, with $\sim$ 3, 3, 3, 2, 2, 2, 2.5, 6, 18, 10, 13, 18, 25, 36$\arcsec$ resolution respectively.
These surveys enabled uniform photometric selection of massive young stellar objects (MYSOs) from point-source catalogs, including the 8$\mu$m catalogs of \citet{gruendl09} and \citet{whitney08} ($\sim$0.5 pc resolution) and the 250$\mu$m catalog of \citet{seale14} ($\sim$2 pc resolution). 
Note that the definition of "massive" YSO is an observational one - all point sources in the catalogs were analyzed with classification cuts in color-magnitude diagrams, then the entire spectral energy distribution was fit with \citet{Robitaille2006} models of single stars with dusty disks and flattened envelopes, to derive a luminosity and mass.  After that analysis, it was determined that the catalogs are fairly complete down to $\sim$5M$_\odot$.
Since 1$\arcsec$=0.25pc at the LMC distance, 2MASS+Spitzer resolution at 2.2, 4.5, 8, and 24$\mu$m of 0.7, 0.5, 0.6, and 1.5~pc permits robust comparison to JWST observations at 1~Mpc distance, with resolution of 0.4, 0.8, 1.1, and 3.1~pc  at 2, 4.5, 10, and 21$\mu$m, respectively.  The combined LMC MYSO catalogs with photometry at those four wavelengths contains 2036 sources.

\subsection{ALMA}

We use Atacama Large Millimeter/submillimeter Array (ALMA) CO(2-1) data from the Morita Atacama Compact Array (ACA) to map the distribution of 
molecular gas in these regions.  These data were observed in projects 
2018.A.00062.S (NGC~300 and NGC~7793), 2017.1.00901.S (M33), and 2018.A.00058.S (M33).  The data from NGC~300 and NGC~7793 include total power data and were reduced and imaged in the PHANGS-ALMA survey \citep{phangs-alma}. The M33 data are only from the interferometer, and are also imaged using the PHANGS-ALMA imaging pipeline \citep{phangs-pipeline}.  In this work, we use the signal-masked integrated intensity maps produced by the pipeline.

\section{Source Detection, Photometry, and Deblending}

\label{sec:photometry}

\begin{table*}
    \centering
    \begin{tabular}{cccccc}
    \hline
    && F200W/F210M & F430M/F444W(F360M) & F1000W & F2100W \\
    \hline
    &Camera & NIRCam-SW & NIRCam-LW & MIRI & MIRI \\
    &Pixel size [arcsec] & 0.031 & 0.063 & 0.11 & 0.11 \\
    &PSF FWHM [arcsec] & 0.066/0.071 & 0.144/0.145(0.12) & 0.33 & 0.67 \\
    \hline
d=0.8Mpc & Physical scale [pc] & 0.293 & 0.60 & 1.35 & 2.78 \\
M33 & integration time [s] & 773 & 773 & 366 & 189 \\
    & sensitivity [$\mu$Jy] & 0.15 & 0.20 & 5.5 & 6.0 \\
    \hline
d=2.0Mpc & Physical scale [pc] & 0.64 & 1.39 & 3.18 & 6.52 \\
NGC300 & integration time [s] & 515 & 515 & 733 & 1443\\
    & sensitivity [$\mu$Jy] & 0.035 & 0.04 & 0.85 & 2.5 \\
    \hline
d=3.7Mpc & Physical scale [pc] & 1.15 & 2.59 & 5.93 & 12.2 \\
NGC7793 & integration time [s] & 687 & 687 & 2153 & 4373 \\
    & sensitivity [$\mu$Jy] & 0.015 & 0.02 & 0.5 & 1.5 \\
   \hline
d=5.2Mpc & Physical scale [pc] & 1.66 & 3.02 & 8.53 & 17.1 \\
NGC5068 & integration time [s] & 1203 & 430 & 122 & 322 \\
    & sensitivity [$\mu$Jy] & .079 & .17 & 1.9 & 8.0\\

    \hline
    \hline
    \end{tabular}
    \label{Tab:bands}
    \caption{
    Properties of the JWST observations used in this work at 2, 4, 10, 21 $\mu$m. Sensitivities provided are from the Exposure Time Calculator, 5$\sigma$ for point sources; in practice for $\lambda\geq$4.5$\mu$m, compact source detection is limited by structured interstellar emission, to $\sim$0.4mJy independent of distance.
    For M33, medium-band filters F210M and F430M were used instead of the broadband filters to mitigate against saturation from bright sources.  
    For NGC~5068, the data are obtained as part of the PHANGS-JWST survey, which did not observe with F444W, 
    so instead we use the filter closest in wavelength F360M. }
    \label{Tab:bands}
\end{table*}

F1000W is used as the principal detection filter because it provides higher angular resolution and sensitivity than F2100W while still tracing warm dust emission from embedded sources. As described below, sources detected at 2$\mu$m and 4$\mu$m are associated with each 10$\mu$m detection, while at 21$\mu$m, we perform forced PSF-photometry at the location of each 10$\mu$m detection (deblending -- see below).   10~$\mu$m sources without counterparts at both 2 and 4~$\mu$m are removed from the sample to reduce the likelihood of spurious detections associated with local peaks in the diffuse ISM, and to facilitate the subsequent analysis which requires detection at all four wavelengths (i.e., 10 $\mu$m for detection; 10 $\mu$m and 21$\mu$m for color selection; 2$\mu$m and 4$\mu$m for visual classification, morphological analysis, and mass estimation).

First, source extraction was performed independently on the 2$\mu$m (F210M/F200W), 4$\mu$m (F430M/F444W), and 10$\mu$m (F1000W) images using the {\sc photutils} \texttt{find\_peaks} algorithm \citep{larry_bradley_2022_6825092}.  
Depending on the detection band, we identified sources between 2–5 sigma above the background level, with a minimum separation between sources of 3–5 pixels. These parameters were chosen to maximize the number of detections, since the catalog is significantly cleaned and filtered in the subsequent steps. 
The background level was estimated using the \texttt{SExtractorBackground} algorithm implemented in \texttt{photutils}, evaluated within square boxes ranging from $10\times10$ to $50\times50$ pixels. This method computes the background using a Source Extractor-like mode estimator based on the image median and mean values.

The 2, 4, and 10$\mu$m source lists were then merged through positional cross-matching. The 4$\mu$m catalog was first matched to the 2$\mu$m catalog using a search radius of 0\farcs144, corresponding to the size of the full width at half maximum of the point spread functions (PSF-FWHM) at 4$\mu$m (Table~\ref{Tab:bands}). The resulting 2$\mu$m and 4$\mu$m source list was then matched to the 10$\mu$m catalog using a search radius of 0\farcs328, corresponding to the PSF-FWHM at 10$\mu$m. In both cases, the nearest positional match was retained for each source. 

Photometry at 2$\mu$m and 4$\mu$m was measured using circular apertures corresponding to the size of the PSF-FWHM in each band (see Table~\ref{Tab:bands}). The background was estimated as the median value measured in an annulus centered on the source, with inner and outer radii set to 2 and 3 times the aperture radius, respectively.

The angular resolution degrades linearly with wavelength, causing emission resolved at 2 and 4 $\mu$m to become significantly blended at 10 and 21$\mu$m. To measure photometry in the 10$\mu$m and 21$\mu$m images (F2100W), we performed fitting using model PSFs generated with {\tt stpsf} \citep{webbpsf} at the locations of the 10$\mu$m detections.  That is, forced PSF-fitting photometry at the location of each 10$\mu$m detection is performed on the 21$\mu$m images, and independent source detection is not performed at 21$\mu$m.

To account for both mosaicking-induced PSF broadening and the possibility that some sources are slightly extended, three PSF models were constructed by convolving the nominal PSF with Gaussians of width $\sigma=1/4$, $3/4$, and $5/4$ pixels at 10$\mu$m.   
In practice, we found that most sources are best fit by the narrowest PSF. Only the central region of the PSF was used in the fitting, corresponding to a radius of 2.7 times the PSF-FWHM (8 pixels for F1000W), which encompasses the central lobe and the first Airy ring.  The PSFs were rotated to match the position angle of the original frames contributing to the mosaic so that the azimuthal PSF structure is correctly aligned. Tests showed that this rotation has only a minor effect on the results.

Sources were fit iteratively starting with the brightest 10$\mu$m detections. For each source, all neighbors within four times the trimmed PSF radius were fit simultaneously along with a local background level. The free parameters in the fit were the source amplitude and the choice of PSF width; the source position was fixed to the location determined by {\sc photutils}. Tests allowing the position to vary by $\pm1$ pixel produced negligible differences in the results. 

The difference between aperture and PSF-fitted flux densities in F1000W is modest across the population: for sources brighter than $2\times10^{-2}$ mJy (the 10~$\mu$m selection threshold adopted as discussed in Section~4.2), the median ratio of PSF to aperture flux density is 0.04 dex (sources are 10\% fainter because those fluxes no longer contain contaminating sources in the aperture), and the standard deviation is 0.08$\;$dex. 
Once the 10$\mu$m sources were fit, we repeated the procedure at 21$\mu$m using the 10$\mu$m source positions and the appropriate PSF model for F2100W.
The deblending of sources at F2100W is more consequential, as expected - most F1000W sources have a contaminating source within the aperture at F2100W resolution, which is why this PSF-fitting deblending is needed.  For sources above 2$\times$10$^{-2}$mJy, the median ratio of PSF-fit to aperture flux density is -0.6 dex, with 0.4 dex RMS.  There are two clear populations - about 3/4 of the sources get deblended, and about 1/4 are isolated, and their flux remains within 0.2 dex of the aperture photometry.

\subsection{Detection and photometry on convolved images}
\label{sec:convolved_images}

To better understand how the photometry of embedded sources changes with increasing distance and decreasing angular resolution, we processed the images to reproduce the resolution and noise as if each nearby galaxy were observed by JWST at the distances of the more distant galaxies in the sample (Table~\ref{tab:distance_effects}).  We generated convolution kernels from each JWST PSF (Table~\ref{Tab:bands}) to a Gaussian PSF at the coarser resolution appropriate for each larger distance and each filter, using jwst\_kernels \footnote{\url{https://github.com/francbelf/jwst_kernels}} \citep{2024Williams,aniano11}. Images convolved to a greater distance still have different resolution in each JWST band, just as if the galaxy were observed at that greater distance. After convolution, each image was binned an integer number of pixels so that the pixels per beam of the convolved images are as close as possible to actual JWST images. Finally, the noise level was measured in both the synthetic image and the corresponding image of the galaxy observed at the larger distance in the same filter, using the sigma-clipped standard deviation (5$\sigma$ clipping, two iterations). Random Gaussian noise was then added to the sigma-clipped synthetic image to bring its noise level up to the same level as the galaxy actually observed at the larger distance (alternately, one can consider this operation to be decreasing the brightness of the sources as 1/distance$^2$ while holding the noise constant). 

We then repeated our procedures for source detection and photometry on the convolved images.

\begin{figure*}
    \centering
    \includegraphics[width=0.23\linewidth]{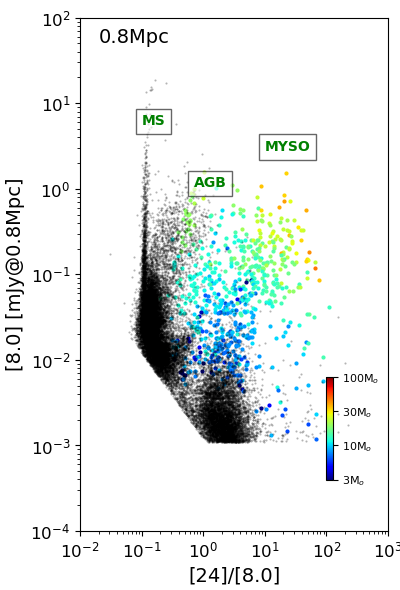}
    \includegraphics[width=0.23\linewidth]{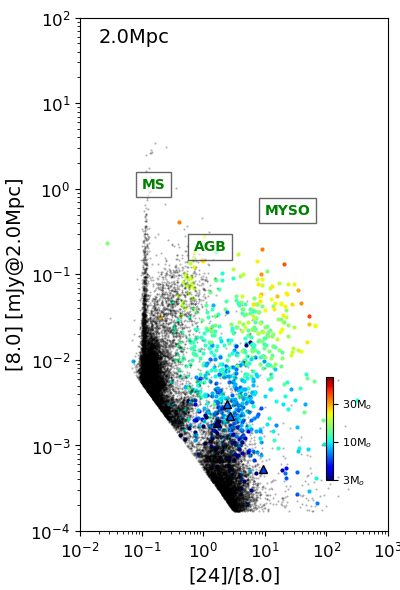}
    \includegraphics[width=0.23\linewidth]{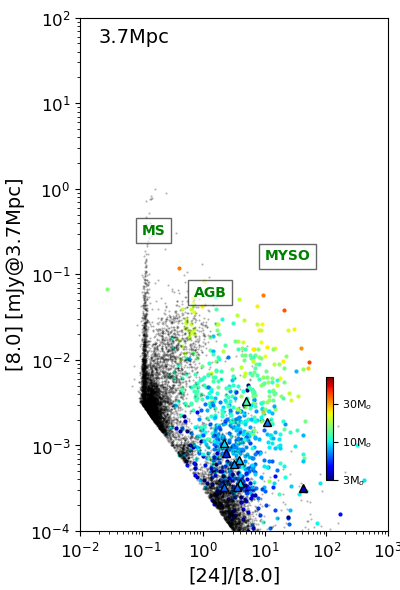}
    \includegraphics[width=0.23\linewidth]{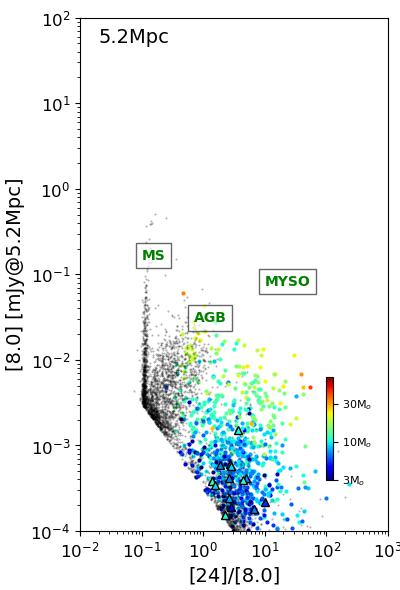}\\
    \includegraphics[width=\linewidth]{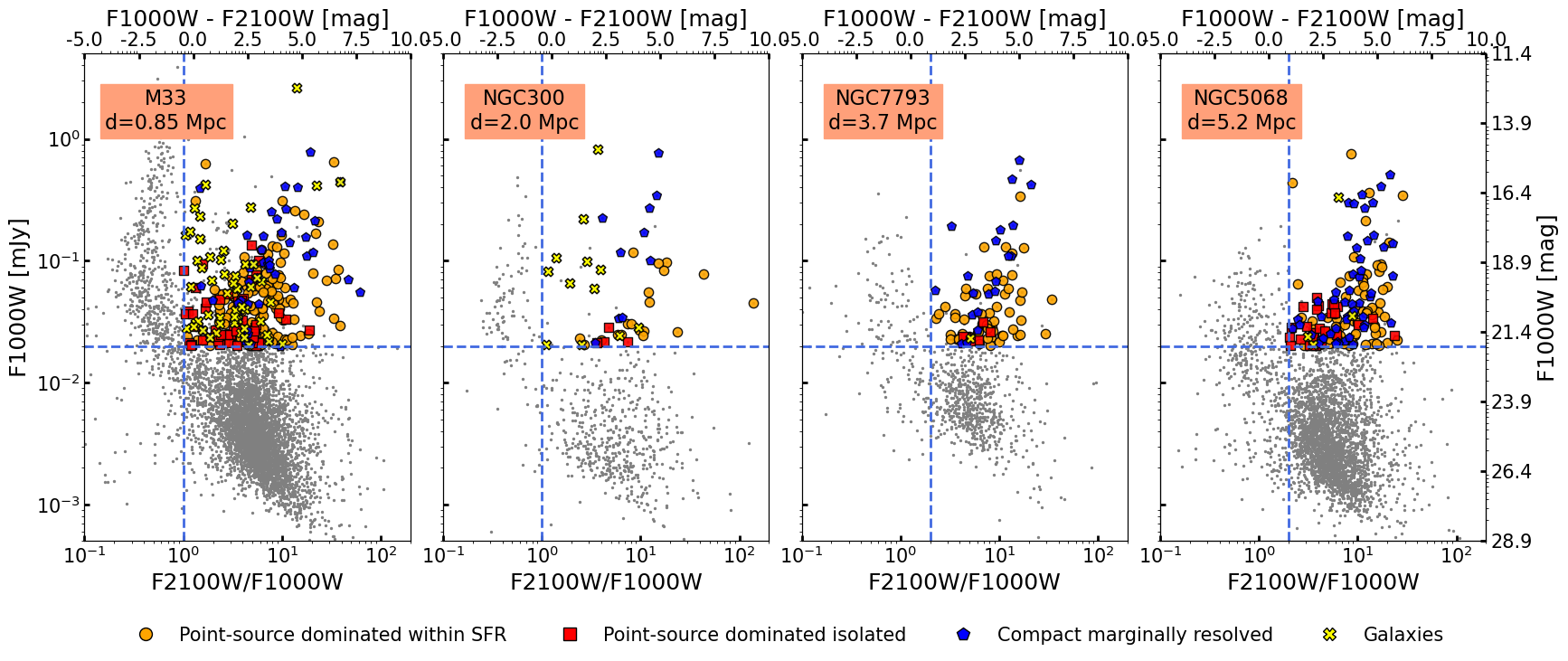}
   \caption{\textit{Top Panels:}
Color-–magnitude diagrams of point sources from the SAGE LMC catalog projected to larger distances.  The diagrams use photometry at 8$\mu$m (IRAC4) and 24$\mu$m (MIPS1). Objects identified as MYSOs \citep{whitney08, gruendl09} are color-coded by stellar mass derived from single-source dust radiative-transfer modeling of their spectral energy distributions. Regions occupied by main-sequence and AGB stars are indicated. The CMDs are simulated as if observed at the distances of M33 (0.8 Mpc), NGC~300 (2.0 Mpc), and NGC~7793 (3.7 Mpc) to study the effects of source blending and inform selection with increasing distance. The colors of MYSOs remain largely unchanged with increasing distance, the small fraction of  MYSOs whose colors change by more than a factor of two as a result of blending are shown with triangles. The features occupied by MS and AGB stars remain distinct, and a mid-infrared color selection is expected to remain robust across the 1–5~Mpc distance range.  
\textit{Bottom panels:} Similar color--magnitude diagrams but using 10 and 21\,$\mu$m MIRI photometry for the galaxies studied in this work. Colors and symbols represent objects in the different categories (see Sec.~\ref{subsec:classification}).}
   \label{fig:lmc_cmd_10_24}
\end{figure*}

\section{Selection of Young, Compact Dusty Objects}
\label{sec:modeled_CMD}

As discussed earlier, MYSOs in the Magellanic Clouds were extensively studied with Spitzer.    
Here we use the SAGE catalogs to inform our selection of point and point-like young dusty sources in more distant galaxies with JWST, using SAGE photometry in the filters closest to the 2, 4, 10, and 21 $\mu$m observations made with JWST: 4.5 (IRAC2), 8 (IRAC4), and 24 $\mu$m (MIPS1) and $K$-band (2MASS, 2.16$\mu$m).  JWST observations of M33 at a distance of 0.85 Mpc achieve a comparable physical resolution to SAGE (0.5–1.5 pc across 1–25$\mu$m), and the colors are very similar (see Appendix~\ref{append:color_corr}), 
which enables the application of similar selection methods.

\subsection{Mid-IR Selection Diagrams Based on SAGE MYSOs}
\label{sec:SAGE_experiments}

The top panel of Figure~\ref{fig:lmc_cmd_10_24} shows the 8 $\mu$m vs.\ 24/8 $\mu$m color–magnitude diagrams (CMDs) as they would appear at the distances of M33 (0.8 Mpc), NGC~300 (2.0 Mpc), NGC~7793 (3.7 Mpc) and NGC~5068 (5.2 Mpc), constructed from photometry in the SAGE LMC point source catalog. In this section we focus on mid-IR color selection using these CMDs. The 4.5 $\mu$m vs.\ 4.5 $\mu$m/$K$ CMDs are presented later alongside the analogous JWST CMDs in the discussion of candidate morphology to facilitate direct comparison (Section \ref{sec:CI}).

To construct the CMDs, we begin with the SAGE LMC point source catalog and project the sources to the distances of M33, NGC~300, NGC~7793 and NGC~5068. That is, for the panels corresponding to 2.0 Mpc (NGC~300), 3.7 Mpc (NGC~7793) and 5.2 Mpc (NGC~5068), the SAGE photometry is first dimmed to the appropriate distance, while the left panel (0.8 Mpc, corresponding to M33) uses the original SAGE photometry.  A sensitivity limit comparable to that of the JWST observations is imposed as listed in Table~\ref{Tab:bands}.
In practice, the threshold for reliable compact source detection is much higher since the background is dominated by structured ISM emission, so these limits do not affect the conclusions here.   

To mimic the angular resolution of JWST at these distances, sources separated by less than the JWST resolution at the relevant wavelength have their fluxes combined.
The small fraction of MYSOs whose colors change by more than a factor of two as a result of blending are shown with triangles (see also next section).
In the left panel, to represent SAGE MYSOs at the distance of M33, the modifications to the original Spitzer data are therefore minimal: only the resolution-based source combination ($<10^{-3}$ of the sources) and the JWST-like sensitivity cutoff are applied.

MYSOs have been shown to be well separated from main-sequence stars and the majority of AGB stars with mid-infrared colors \citep{sage,2007Bolatto}, and this is clearly illustrated in Figure~\ref{fig:lmc_cmd_10_24} (top panels).  The separation of the distinct features persists in the CMDs projected to a distance of 5.2 Mpc. The areas where main sequence and AGB stars are found are indicated, while objects classified as MYSOs are color-coded according to mass \citep[computed through radiative transfer modeling of photometry covering the full optical to far-IR spectral energy distribution, e.g.,][]{whitney08,gruendl09,seale14}.   


In most cases, the 24/8 $\mu$m colors of MYSOs do not change significantly with increasing distance, although sources identified at high spatial resolution may merge with neighboring point sources. While this may seem trivial, it is important to verify that when MYSOs blend with other objects—typically bluer sources—the resulting mixed source remains dominated by the MYSO in the mid-infrared over this range of resolutions. Blending between two MYSOs with similar colors will, of course, produce a source with similar mid-infrared colors. \emph{Therefore, a mid-infrared color selection is expected to remain robust across the 1–5~Mpc distance range.} The next question is whether the properties of individual MYSOs can still be recovered at these distances, or whether blending limits the analysis to integrated cluster emission.

It should be noted however that a population of ``extreme'' AGB stars, with mass-loss rates $\gtrsim$ 10$^{-6}$ -- 10$^{-4}$ M$_\odot\;$yr$^{-1}$, can overlap with MYSOs in infrared colors \citep{2025FrASS..1287415L,Boyer2011,vanloon06}. 
To minimize contamination from these stars when detecting YSOs in NGC~6822 ($\sim$0.5 Mpc) with JWST, \citet{lenkic24} applied a magnitude cut at the bright end.  As discussed in the Spitzer-based literature referenced earlier, these infrared-bright sources can be difficult to distinguish as evolved or young stars, especially without spectroscopy.  In the current study, our goal is to study the full population of young dusty point and point-like sources, including star clusters, 
so we keep such bright red sources in our catalog and analysis, acknowledging that a few may not turn out to be associated with star formation once more data are obtained. (The small numbers of sources, and other evidence in favor of keeping them, are provided below in \S~\ref{sec:CI}.)

\subsubsection{Can individual MYSOs be studied in galaxies beyond the MCs with JWST?}
\label{sec:myso_merge}

\begin{figure}
    \centering
    \includegraphics[width=\columnwidth]{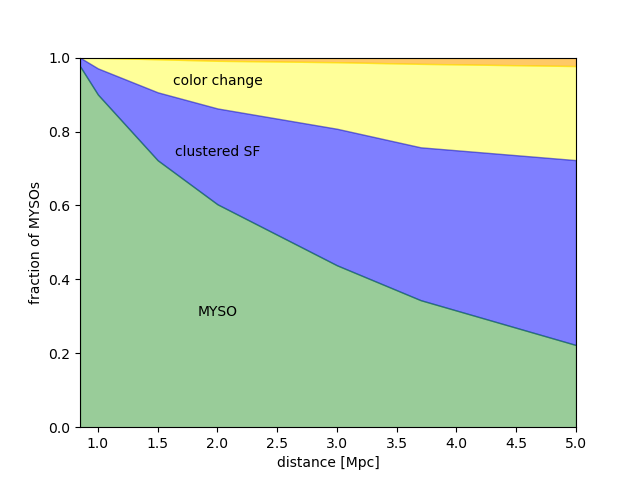}
    \caption{Predictions, based on Spitzer SAGE-LMC data, of the impact of source blending in JWST imaging as a function of distance on the characterization of dusty embedded star clusters. Green: fraction of clusters dominated by a single MYSO whose infrared colors change by less than 0.3~dex as the aperture size increases to simulate larger distances. In this regime the properties of individual MYSOs (e.g., masses and ages) can be reliably characterized. This fraction drops below 50\% at distances of $\sim$3~Mpc. Blue: clusters containing multiple MYSOs where blending does not change the infrared colors by more than 0.3~dex. These objects would still be identified as dusty star formation, but the derived properties correspond to the combined emission of multiple sources. Yellow: cases where blending causes infrared color changes greater than 0.3~dex.  These will still be identified as compact dusty star formation, but again, derived physical parameters may not reliably represent only the most massive source in the group. Finally, the narrow orange wedge at the top represents sources that become too blue to be identified as dusty star forming objects from its mid-infrared colors. 
    \label{myso_merge}}
\end{figure}

While the CMD analysis above shows that mid-infrared colors remain a robust way to identify dusty star-forming sources to distances of at least several Mpc, an additional question is whether individual MYSOs can still be characterized at these distances.

At the $\sim$1~pc resolution of the \textit{Spitzer} SAGE survey, a point source is unlikely to correspond to a single YSO, but rather to a small embedded cluster. This was shown in follow-up observations of SAGE sources with HST, JWST, and ground-based adaptive optics \citep[e.g.,][]{Chen10, Stephens17}. 

Nevertheless, 
\citet{Chen10} showed that the uncertainties introduced by modeling such sources as single MYSOs are relatively modest ($\sim$20--40\% in derived properties such as photospheric mass, envelope mass, and disk mass) and do not significantly affect conclusions about star-formation timescales, efficiencies, or star-formation laws derived from MYSO counts.

As described above, we merged SAGE MYSOs with all point sources in the SAGE catalog to investigate how the colors and magnitudes of those sources change with simulated distance.
The results of the analysis are presented in Figure~\ref{myso_merge} and indicate that studies of compact dusty star formation analogous to the SAGE MYSO analyses should remain feasible out to $\sim$3Mpc.  Between 3 and 4 Mpc, the fraction of MYSOs with significant color changes increases above 50\%.

\subsection{Mid-IR Candidate Selection Criteria}
\label{sec:classification}
\begin{table*}[]
\centering
\scriptsize
\begin{tabular}{cccccccc}
\hline
Galaxy & Selected in 21/10 &
\shortstack{Point-source\\within SFR} &
\shortstack{Point-source\\isolated} &
\shortstack{Marginally\\resolved} &
\shortstack{Background\\galaxies} &
\shortstack{Massive\\stars} &
Other* \\
\hline
M~33 & 415 & 135 (32.5\%) & 31 (7.5\%)& 50 (12\%) & 55 (13.3\%) & 36 (8.7\%) & 108 (26\%)\\
NGC~300 & 52  & 18 (34.6\%)  & 4  (7.7\%) & 10 (19.2\%) & 12 (23\%) & 1 (2\%) & 7 (13.5\%)\\
NGC~7793 & 86 & 54 (62.8\%)& 6 (7\%) & 20 (23.3\%)& 1 (1.1\%) & 4 (4.6\%) & 1 (1.1\%)\\
NGC~5068 & 162 & 66  (40.7\%)& 23 (14.2\%) & 50 (31\%)& 4 (2.5\%) & 14 (8.6\%) & 5 (3\%)\\
\hline
\hline
\end{tabular}
\caption{Number of candidates selected in the 10 vs.\ 21/10 CMD (Col.~2), subdivided into three visually defined categories of young dusty compact sources (Cols.~3–5), background galaxies (Col.~6), massive stars (col~7) and other interlopers (Col.~8). 
Percentages indicate the fraction of the total number of mid-IR CMD selected sources in each galaxy (Col.~2).  
\textasteriskcentered~Other = Evolved Stars, artifacts, or unidentified sources.} 
\label{Tab:categories}
\end{table*}

We construct JWST F1000W vs. F2100W/F1000W CMDs for each of the galaxies in our sample and compare them to the SAGE LMC 8~$\mu$m vs.~$24/8,\mu$m CMDs presented in the previous section. These CMDs are presented in the lower panels of Fig.~\ref{fig:lmc_cmd_10_24}.  

The JWST CMDs show a blue branch similar to that formed by main-sequence stars in the SAGE diagrams, although with greater scatter. An adjacent branch to the red, likely corresponding to AGB stars, can also be distinguished. Finally, a broad distribution of even redder sources is present, where MYSOs are expected to lie.  Of course, as the distance of the host galaxy increases, only the brightest stars and MYSOs can be detected, and source density of candidates in each of these CMD features should decrease.




Based on these CMDs, we adopt the following color criteria to separate MYSOs from main sequence and AGB stars:
\begin{itemize}
    \item $\rm F2100W/F1000W > 1$ for M33 and NGC~300,
    \item $\rm F2100W/F1000W > 2$ for NGC~7793 and NGC~5068.
\end{itemize}
The blue branch appears to shift toward redder colors with increasing distance due to source blending; therefore, we adopt a higher color cut for the more distant galaxies.  We also apply a flux limit of 
\begin{itemize}
    \item $\rm F1000W = 2\times10^{-2}$ mJy for all galaxies.
\end{itemize}
The adopted F1000W limit corresponds to the magnitude where the separation between the blue and red branches of the CMD becomes less distinct, and to a $\sim50\sigma$ point source detection. Comparison with the SAGE MYSO CMDs (Fig~\ref{fig:lmc_cmd_10_24}, top panels) at a Spitzer 8 $\mu$m flux of 0.02 mJy provides a first rough estimate of the MYSO mass range detectable with our JWST observations. We return to this in Section~\ref{sec:masses} when we independently estimate these masses.

These selection criteria yield a total of 715 candidates: 415, 52, 86 and 162 in M33, NGC~300, NGC~7793 and NGC~5068 respectively (Table~\ref{tab:sample}).   

\begin{sidewaystable*}
\centering
\vspace*{-7cm}
\scriptsize
\begin{tabular}{cccccccccccccccccc}
\hline
Galaxy & ID & RA & DEC  & RA & DEC  & RA & DEC & Flux & Err & Flux & Err & Flux & Err & Flux & Err & CI & Label\\
  & & 10$\mu m$ & 10$\mu m$  & 4$\mu m$  & 4$\mu m$  & 2$\mu m$  & 2$\mu m$  & 21$\mu m$  & 21$\mu m$  & 10$\mu m$  & 10$\mu m$  & 4$\mu m$  & 4$\mu m$  & 2$\mu m$  & 2$\mu m$ & & \\
  & & [deg] & [deg] & [deg] & [deg] & [deg] & [deg] & [mJy] & [mJy] & [mJy] & [mJy] & [mJy] & [mJy] & [mJy] & [mJy] & & \\
\hline
M33 & 0 & 23.48930 & 30.54491 & 23.48922 & 30.54490 & 23.48921 & 30.54489 & 2.013e-01 & 3.289e-02 & 2.546e-02 & 4.160e-03 & 1.963e-04 & 6.851e-05 & 2.080e-04 & 1.208e-05 & 2.03 & PSD-SFR \\
M33 & 3 & 23.40119 & 30.55301 & 23.40111 & 30.55303 & 23.40110 & 30.55303 & 1.459e-01 & 4.475e-02 & 2.690e-02 & 1.229e-02 & 1.368e-04 & 1.159e-04 & 1.185e-03 & 1.782e-05 & 1.51 & PSD-SFR \\
M33 & 5 & 23.37471 & 30.53048 & 23.37480 & 30.53051 & 23.37481 & 30.53051 & 2.478e+00 & 1.610e+00 & 7.140e-02 & 5.698e-02 & 1.784e-04 & 1.136e-04 & 1.471e-03 & 1.931e-05 & 1.76 & PSD-SFR \\
M33 & 6 & 23.49440 & 30.57100 & 23.49446 & 30.57095 & 23.49446 & 30.57094 & 9.505e-02 & 1.724e-02 & 2.018e-02 & 4.359e-03 & 1.547e-04 & 2.491e-05 & 1.197e-03 & 1.434e-05 & 1.49 & PSD-SFR \\
M33 & 10 & 23.50296 & 30.58461 & 23.50296 & 30.58461 & 23.50295 & 30.58461 & 9.874e-02 & 2.307e-02 & 2.557e-02 & 6.081e-03 & 5.388e-04 & 4.231e-05 & 3.022e-04 & 1.749e-05 & 1.96 & PSD-SFR \\
M33 & 11 & 23.45895 & 30.56186 & 23.45902 & 30.56190 & 23.45903 & 30.56190 & 1.267e-01 & 6.324e-02 & 2.356e-02 & 8.838e-03 & 1.326e-04 & 3.871e-05 & 1.026e-03 & 1.625e-05 & 1.54 & PSD-SFR \\
M33 & 17 & 23.49908 & 30.57734 & 23.49904 & 30.57737 & 23.49904 & 30.57737 & 2.083e-01 & 5.978e-02 & 4.190e-02 & 9.987e-03 & 1.492e-04 & 6.780e-05 & 9.508e-04 & 1.766e-05 & 1.61 & PSD-SFR \\
M33 & 21 & 23.32813 & 30.55565 & 23.32812 & 30.55571 & 23.32810 & 30.55571 & 1.601e-01 & 4.784e-02 & 2.985e-02 & 7.512e-03 & 4.615e-05 & 7.584e-05 & 3.129e-04 & 2.138e-05 & 1.78 & PSD-SFR \\
M33 & 22 & 23.44862 & 30.55114 & 23.44864 & 30.55115 & 23.44864 & 30.55115 & 5.277e-01 & 9.133e-01 & 6.506e-02 & 2.251e-02 & 8.724e-03 & 3.378e-04 & 5.725e-02 & 1.386e-04 & 1.29 & Star \\
M33 & 24 & 23.34949 & 30.53149 & 23.34948 & 30.53153 & 23.34948 & 30.53154 & 1.797e-01 & 2.655e-02 & 2.093e-02 & 8.643e-03 & 3.042e-05 & 4.198e-05 & 2.131e-04 & 1.474e-05 & 1.99 & PSD-SFR \\
\hline
\hline
\end{tabular}
\caption{
Catalog of the 715 sources selected based on their mid-infrared colors, i.e. $F_{\rm 2100W}/F_{\rm 1000W} > 1$ for M33 and NGC~300, and $> 2$ for NGC~7793 and NGC~5068, together with a flux threshold of $F_{\rm 1000W} > 2\times10^{-2}$ mJy. Coordinates correspond to the centroid measured in each detection band. The catalog includes photometric measurements in the four bands, concentration index values, and classification labels indicating whether each source is classified as a point-source dominated source within star-forming regions (PSD-SFR), an isolated point-source dominated source (PSD-ISO), or a compact marginally resolved source (MRCL), as well as contaminants (e.g., Stars, galaxies (Gal), and diffraction artifacts (DA)). The full catalog is available online.} 
\label{tab:sample}
\end{sidewaystable*}



\subsection{Visual Classification \& Contaminant Removal}
\label{subsec:classification}


MJR visually inspected each of the 715 objects (with spot checks by RI and BCW) across all available UV–optical HST and infrared JWST bands. The inspection identified contaminants (N=193), which were found to generally consist of background galaxies (Fig.\ref{fig:GAL_M33}), some red evolved stars, and artifacts e.g., from diffraction spikes.  We found that the rest (N=522) could generally be grouped into three candidate categories: Point-source dominated within star forming regions, Point-source dominated isolated sources, and Compact marginally resolved sources.               

These visual categories (defined below) are intended to be generally consistent with the regimes identified in the blending analysis of SAGE-LMC sources presented in Section~\ref{sec:myso_merge} and Figure~\ref{myso_merge}. Sources classified as point-source dominated, either within star-forming regions or in relative isolation,
are expected to correspond to systems in which the mid-infrared emission is dominated by a single massive YSO. In contrast, objects classified as compact marginally resolved
likely correspond to systems where the infrared emission arises from multiple embedded sources within a small region, such that the derived photometric properties represent the combined emission of several objects rather than an individual dominant MYSO. We present these classifications as an exploratory way of separating candidate sources into regimes that roughly correspond to the single-source and blended-source cases, and compare them with measurements of the concentration index at 2$\mu$m in the next section.

We relied heavily on the 2$\mu$m images which provide superb resolution to visually check whether the 21/10 $\mu$m selected 10 $\mu$m detections were point sources or slightly more extended, following well-established procedures for inspection of star cluster candidates in HST observations \citep[e.g.,][]{adamo17, deger22}. The primary distinction between categories is whether the source appears point-like or slightly extended at 2$\mu$m; the subdivision of 
point-source dominated into “isolated” and “within star-forming regions” reflects differences in their local environments.  The characteristics of each category are as follows, and examples of objects in each are presented in the Appendix~\ref{Ap:fig_examples}  (Figures~\ref{fig:YSO_SFR}-~\ref{fig:CLTs}).

\textbf{Point-source dominated sources within SFRs}
(Fig.~\ref{fig:YSO_SFR}): 
The candidates appear as point sources at 2$\mu$m and in some of the HST bands, and are within associations, or even more extended star-forming regions, where individual members can be distinguished. In these cases, the objects are surrounded by diffuse ISM, and often accompanied by nearby optically blue stars, suggesting an actively star-forming environment.  

\textbf{Point-source dominated sources isolated} 
(Fig.~\ref{fig:YSO_isolated}):
These candidates are similar to the previous category, but are relatively isolated, with no obvious stellar association, or other young stars nearby.  However, they appear surrounded by diffuse ISM, and lie along ISM filaments (as indicated by dust emission in the JWST MIRI images and/or a dust lane or patch in the HST images).  Sometimes they appear to be located at the edges of cavities or bubbles (e.g. M~33 ID: 139; NGC~7793 ID: 4, NGC~5068 ID: 42, 52, 108 Fig.~\ref{fig:YSO_isolated}). 



\textbf{Compact marginally resolved sources}
(Fig.~\ref{fig:CLTs}):
In these cases, the object is slightly more extended than a point source at 2$\mu$m. Sometimes it appears to be part of a larger star-forming region (e.g M~33 ID: 350, 382; NGC~300 ID:13,32, 39; NGC~7793 ID: 30, 84; NGC~5068 ID: 24, 119, 149, 150 Fig.~\ref{fig:CLTs}), while in others it appears more isolated (e.g. M~33 ID: 165; NGC~300 ID: 16, 21, 22 26; NGC~7793 ID: 68). However, in the majority of the cases, the object is either surrounded by or connected to a nearby star-forming region through ISM filaments clearly visible in 10$\mu$m and/or 21$\mu$m.

Note that the separation between these categories is not always clear, and for some objects the classification is ambiguous. 
Some ambiguity is expected given that the blending analysis in Figure~\ref{myso_merge} predicts a gradual transition as a function of galaxy distance between regimes where individual MYSOs dominate the infrared emission and regimes where multiple embedded sources contribute significantly.
The four bottom CMDs in Fig.~\ref{fig:lmc_cmd_10_24} show the different categories with distinct symbols and colors.




\subsection{Near-IR Color Selection Based on SAGE MYSOs \& Concentration Index}
\label{sec:CI} 

\begin{figure*}
    \centering
    \includegraphics[width=\linewidth]{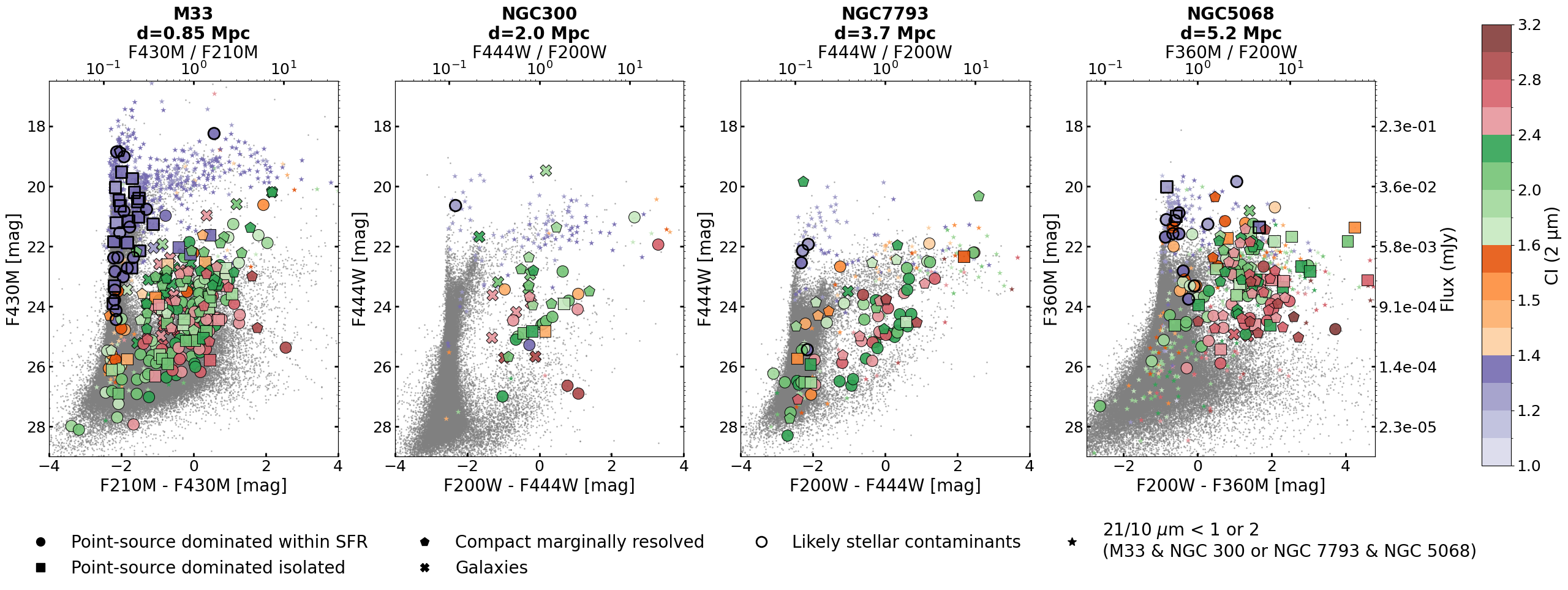} 
    \includegraphics[width=0.23\linewidth]{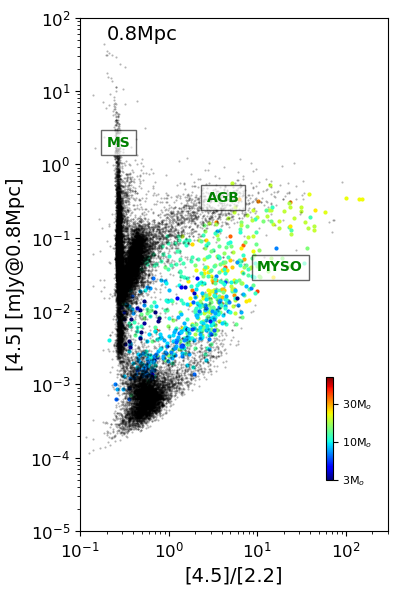}
    \includegraphics[width=0.23\linewidth]{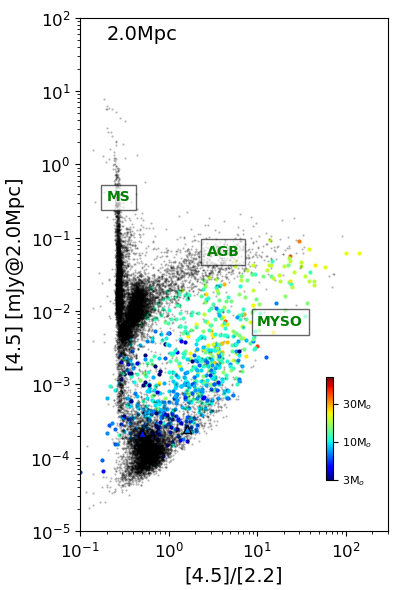}
    \includegraphics[width=0.23\linewidth]{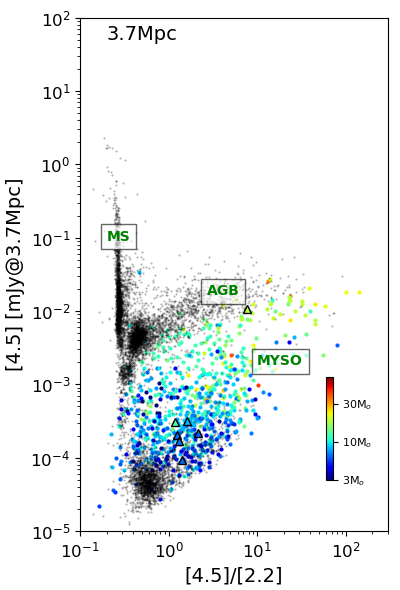}
    \includegraphics[width=0.23\linewidth]{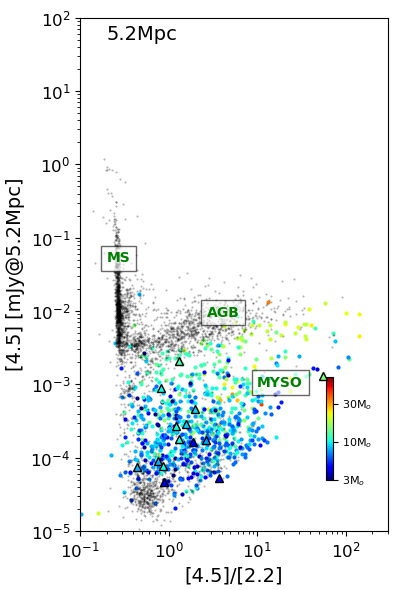}
    \caption{
Near-IR color--magnitude diagrams for the JWST observations presented in this paper (top, for M33, NGC~300, NGC~7793, and NGC~5068), and based on SAGE-LMC (bottom, projected to distances of 0.8, 2.0, and 3.7~Mpc). These diagrams are analogous to the $21/10\,\mu$m (JWST) and $24/8\,\mu$m (Spitzer) CMDs presented earlier in Figures~\ref{fig:lmc_cmd_10_24} and ~\ref{fig:CMD_categories_convolved}, but the JWST CMDs are now color-coded by concentration index (CI). Purple hues correspond to compact sources with CI values typical of stars, while green and red indicate increasingly extended sources. Sources with CI values consistent with stars that lie along the main sequence or AGB branch (black borders) are excluded from our sample and further analysis. In the SAGE panels, symbols are color-coded by stellar mass derived from radiative-transfer modeling of the spectral energy distributions of the MYSOs as in Figures~\ref{fig:lmc_cmd_10_24}. Triangles indicate MYSOs whose colors change by more than a factor of two as a result of blending.}
    \label{fig:CMD_CI}
\end{figure*}    

The concentration index (CI) is commonly used in studies of unresolved star clusters in nearby galaxies ($\lesssim$20 Mpc) to distinguish stars from marginally resolved clusters by measuring the compactness of the light distribution \citep[e.g.,][]{chandar10b,whitmore14,adamo17,deger22,Thilker22}. Here, CI measurements provide a quantitative complement to the exploratory visual classifications described above, and allow us to assess whether the categories exhibit corresponding differences in compactness.

We define the CI as the difference between magnitudes measured within apertures of 1 and 4 pixels at $2\,\mu$m (CI = $m_{1pix}-m_{4pix}$), following the procedure described by \citet{whitmore23}. At the distances of the galaxies in our sample, these apertures correspond to physical scales of approximately 0.13--0.5~pc in M33, 0.3--1.2~pc in NGC~300, 0.5--2~pc in NGC~7793, and 0.8--3~pc in NGC~5068. The $2\,\mu$m band was chosen because it provides the highest spatial resolution among the available bands while primarily tracing stellar continuum emission. Previous work has shown that CI measurements at this wavelength provide a reliable means of distinguishing stars from compact clusters in nearby galaxies \citep{whitmore23,Rodriguez23,Rodriguez25}.

We use the $4\,\mu$m versus $2$--$4\,\mu$m CMDs (Figure~\ref{fig:CMD_CI}) to illustrate how the CI values vary across: 
\begin{itemize} 
\item the three visually defined categories of the dusty young population (point-source dominated within star-forming regions, point-source dominated isolated, and compact marginally resolved sources) for 21/10 $\mu$m color-selected sources,
\item 21/10 $\mu$m color-selected background galaxies removed from the candidate list,
\item objects that lie blueward of the mid-infrared 21/10 $\mu$m color selection.
\end{itemize}
    
For comparison, the bottom row of Figure~\ref{fig:CMD_CI} presents analogous CMDs based on the Spitzer SAGE-LMC catalog, which provide a useful reference for interpreting the JWST diagrams. As in Figure~\ref{fig:lmc_cmd_10_24}, SAGE MYSOs identified through radiative-transfer modeling of their spectral energy distributions are color-coded by mass. In the SAGE CMDs, MYSOs occupy a well-defined region toward red $4.5\,\mu$m/$K$ colors that is distinct from the tight loci of main-sequence and AGB stars. The location of these populations provides a guide to where deeply embedded dusty sources are expected to lie in the JWST CMDs.

MYSOs are not necessarily expected to appear strictly point-like at $2\,\mu$m. Circumstellar envelopes, scattered light, and unresolved multiple systems can broaden the observed light distribution relative to that of an isolated star. As discussed above, the JWST point-source dominated candidates likely correspond to a blend of multiple objects 
whose near-infrared emission is dominated by one or a few luminous embedded objects. Such systems are therefore expected to exhibit a range of CI values extending from those typical of stars to values characteristic of marginally resolved clusters, depending on the relative brightness of the dominant MYSO. Background galaxies selected by the mid-infrared color criteria can also appear extended in the near-infrared and therefore exhibit elevated CI values.

The CI distributions broadly support these expectations. Objects located in the red region of the near-infrared CMD—corresponding to the dusty population identified through the mid-infrared color selection—tend to exhibit larger CI values, indicating that many are marginally resolved at $2\,\mu$m. These objects largely correspond to the visually classified compact marginally resolved and point-source dominated within star-forming region sources. 
In contrast, sources along the blue branch of the CMD, consistent with stellar photospheres, predominantly exhibit low CI values typical of unresolved stars (purple).

A small subset of objects selected by the $21/10\,\mu$m color criteria nevertheless exhibits CI values consistent with point sources. The fraction of objects in the dusty young population with JWST F200W CI$<1.4$ (consistent with unresolved stellar sources \citep{Rodriguez25}) is $40/252$ ($\sim16\%$) in M33, $2/33$ ($\sim6\%$) in NGC~300, $4/84$ ($\sim5\%$) in NGC~7793, and $14/153$ ($\sim9\%$) in NGC~5068. In the SAGE studies, candidate MYSOs identified through infrared colors were subsequently vetted through multi-wavelength spectral energy distribution (SED) modeling, and sources consistent with bare or nearly bare stellar photospheres were removed from the final catalogs \citep{whitney08,carlson12}. 
In the present analysis, the limited IR data available make it difficult to perform SED fitting with dust radiative modeling. We therefore exclude sources with CI $<$ 1.4 that are consistent with the main sequence or AGB branch in the $2$–$4\,\mu$m CMD (objects with thick black borders in Fig.~\ref{fig:CMD_CI}, top panel). These compact sources likely correspond to massive stars with near-photospheric $2$–$4\,\mu$m colors that are nevertheless associated with nearby warm dust, producing the observed red 10-20$\mu$m colors. Many of these objects fall within the visually classified category of point-source dominated isolated. Applying this last criterion, 36, 1, 4, and 14 objects were removed from the samples in M33, NGC~300, NGC~7793, and NGC~5068, respectively.

The CI measurements provide an independent test of the visually defined categories. Sources classified as compact dusty marginally resolved generally exhibit the largest CI values. 
Objects identified as point-source dominated within star-forming regions show intermediate CI values, suggesting that many correspond to small clusters or associations dominated by one or a few luminous embedded stars. In contrast, isolated point-source dominated candidates more frequently exhibit CI values consistent with point sources, indicating that their near-infrared emission is dominated by a single object. The overlap in CI distributions among the categories suggests that these classifications represent a continuum of source structures rather than strictly distinct populations.

In summary, the combination of near- and mid-infrared colors together with CI measurements suggests the presence of three populations in our CMDs: (1) deeply embedded dusty sources, which are red in both $2$--$4\,\mu$m and $10$--$20\,\mu$m colors and often exhibit elevated CI values; (2) massive stars with near-photospheric $2$--$4\,\mu$m colors but red $10$--$20\,\mu$m colors, likely due to heating of nearby dust; and (3) extreme AGB stars, which tend to be red in $2$--$4\,\mu$m but relatively compact and bluer in $10$--$20\,\mu$m. For the purposes of this study we only retain objects consistent with populations (1), 
which trace recent star formation. 
Objects consistent with extreme AGB stars are largely excluded by the adopted color selection, although a small number of ambiguous cases remain. These sources are few in number and relatively faint at $10$ and $21\,\mu$m, and therefore do not significantly affect the results.


\subsection{Final Candidate Sample}
\label{sec:final_sample}

 Following the criteria described in the previous subsections, we constructed
our final sample of dusty young objects using a combination of mid-IR color selection, CI and near-IR constraints, and visual inspection.

We first applied an initial mid-IR selection based on the $\rm F2100W/F1000W$ flux ratio ($\S$~\ref{sec:classification}):
\begin{itemize}
    \item $\rm F2100W/F1000W > 1$ for M33 and NGC~300,
    \item $\rm F2100W/F1000W > 2$ for NGC~7793 and NGC~5068.
    \item  $\rm F1000W > 2\times10^{-2}$\,mJy for all galaxies.
\end{itemize}

To reduce additional contamination from massive stars with dusty environment ($\S$~\ref{sec:CI}), 
we excluded sources that simultaneously satisfy: 
\begin{itemize}
\item  CI $<$~1.4 
\item $2$--$4\,\mu$m colors consistent with main-sequence or AGB stars .
\end{itemize} 

Finally, all candidates were visually inspected $\S$~\ref{subsec:classification}. During this process, obvious background galaxies, imaging artifacts, and evident evolved stars were removed from the sample. Based on their morphology and local environment, the remaining sources were classified into three categories:
\begin{itemize}
    \item point-source dominated sources within star-forming regions,
    \item isolated point-source dominated sources,
    \item compact marginally resolved sources.
\end{itemize}

\begin{figure}
    \centering
    \includegraphics[width=\columnwidth]{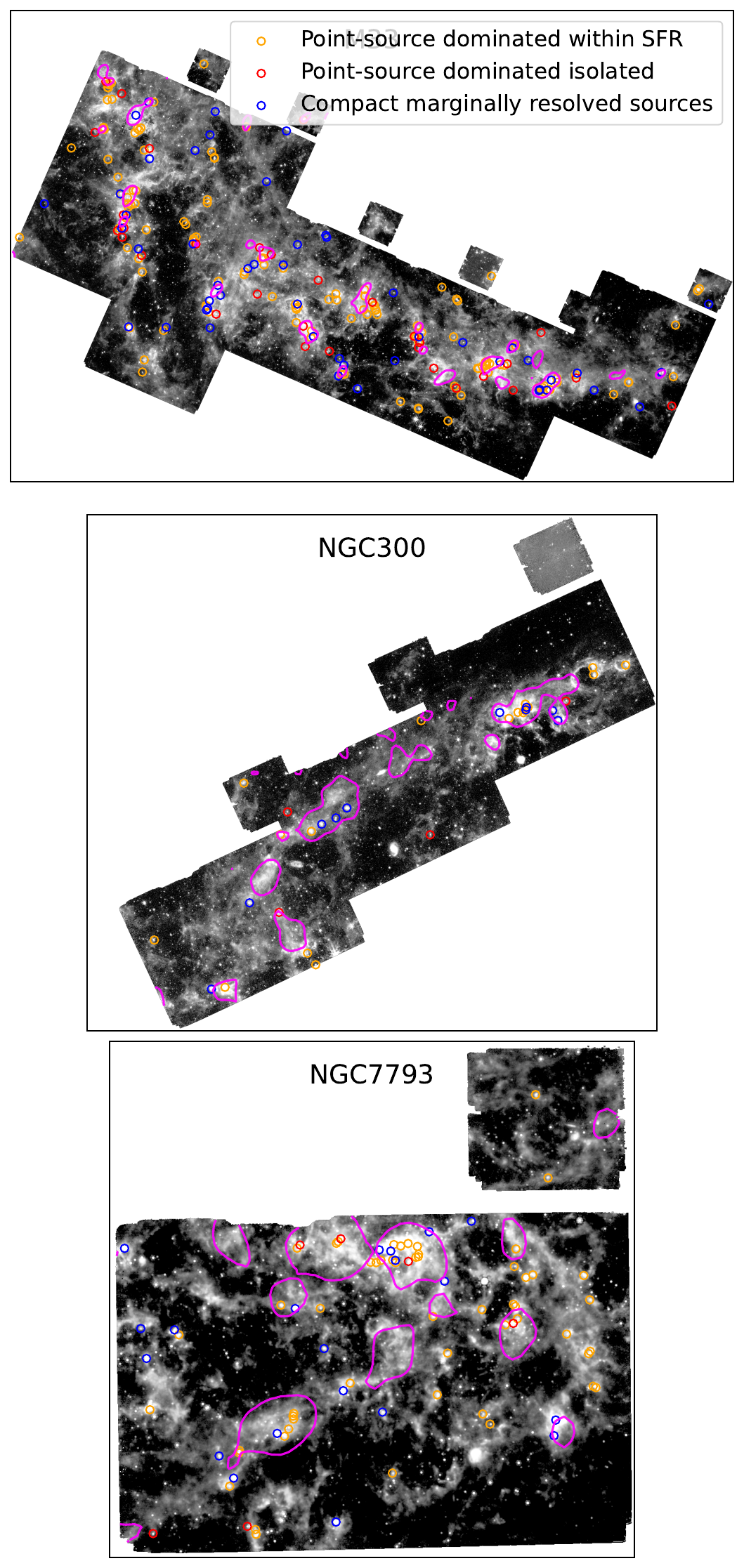}
    \caption{Spatial distribution of the dusty young population overlaid on the JWST F1000W image of each galaxy. Colored circles indicate objects in different categories, while the magenta contours show the 90th percentile of the ALMA CO(2–1) emission.}
    \label{fig:spatial_distribution}
\end{figure}

Summed over these three categories, there are 216, 32, 80, and 139 objects in M~33, NGC~300, NGC~7793, and NGC~5068, respectively, which we refer to collectively as the “dusty young population" hereafter.  This represents 52\%, 62\%, 93\%, and 86\% of the initial candidate sample.
The number of objects in each category is listed in Table~\ref{Tab:categories}.  
In Fig.~\ref{fig:spatial_distribution}, we present the spatial distribution of the ensemble dusty young population 
overlaid on the F1000W images together with ALMA CO(2–1) contours.

Table~\ref{tab:sample} presents the complete sample of 715 sources identified through the mid-IR color selection across the four galaxies. It includes the photometric measurements in the four infrared bands, the corresponding CI values, and classification labels indicating whether each source belongs to one of the dusty young object categories or to contaminant populations such as galaxies, stars, or diffraction artifacts. The full version of the table is available online.

\section{Effects of Distance on Selection}
\label{sec:dusty_pop_at_convolved_images}

Distance affects not only the detectability of individual compact dusty sources but also the interpretation of diffuse mid-infrared emission. A long-standing question in studies of star formation is how much of the mid-infrared emission attributed to diffuse dust heating actually originates from unresolved compact star-forming regions rather than dust heated by older stellar populations. In this section we therefore address two related questions:
\begin{itemize}
\item how increasing distance affects the recovery and selection of compact dusty sources, and
\item how the fraction of mid-infrared emission attributed to compact star-forming regions changes as spatial resolution degrades.
\end{itemize}



\subsection{Recoverability of Individual Dusty Sources}

Using images convolved to simulate observations at larger distances (as described in Section~\ref{sec:convolved_images}), we quantify how many sources are recovered, how many are lost due to source merging or surface-brightness dilution, and how these effects impact the color–magnitude diagram (CMD) selection used to identify dusty young objects. We also assess whether the mid-infrared colors used in our selection remain stable as spatial resolution decreases (as anticipated from the SAGE simulation discussed in Section~\ref{sec:modeled_CMD}), and whether new sources enter the selected CMD region as a result of blending or photometric changes.

We first cross-matched the positions of sources detected at the native 10~$\mu$m resolution with the catalogs obtained from the images which were convolved to simulate observations of galaxies at larger distances (Sect.~\ref{sec:convolved_images}).  The size of the search radius was the size of the convolved FWHM-PSF in each case (i.e., 0\farcs77, 1\farcs39 and 2" for M~33 at 2, 3.6 and 5.2 Mpc respectively, 0\farcs59 and 0\farcs85 for NGC~300 at 3.6 and 5.2~Mpc respectively, and 0\farcs47 for NGC~7793 at 5.2~Mpc).

The recovery fraction decreases steadily with increasing distance (Table~\ref{tab:distance_effects}). In M33, 15\% of the dusty young population is lost when the image is convolved to 2~Mpc, increasing to 25\% at 3.6~Mpc and 46\% at 5.2~Mpc. In NGC~300 the losses are smaller, amounting to 6\% and 9\% at 3.6 and 5.2~Mpc, respectively. In NGC~7793, 28\% of the objects detected at the native 3.6~Mpc resolution are lost when the image is convolved to 5.2~Mpc. These results are consistent with the LMC blending experiments (Sect.~\ref{sec:myso_merge}) in which half of MYSOs blend into other objects by a simulated distance of 3~Mpc, and are ``lost'' as individual sources (although retained as compact dusty star formation regions)

\subsubsection{Visual inspection of lost sources}

To determine the reason for these losses, MJR visually inspected each object not recovered in the convolved images (with spot-checks by RI) to assess the relative contributions of source merging and surface-brightness dilution into the diffuse ISM.

\begin{figure}
    \centering
      \includegraphics[width=1\columnwidth]{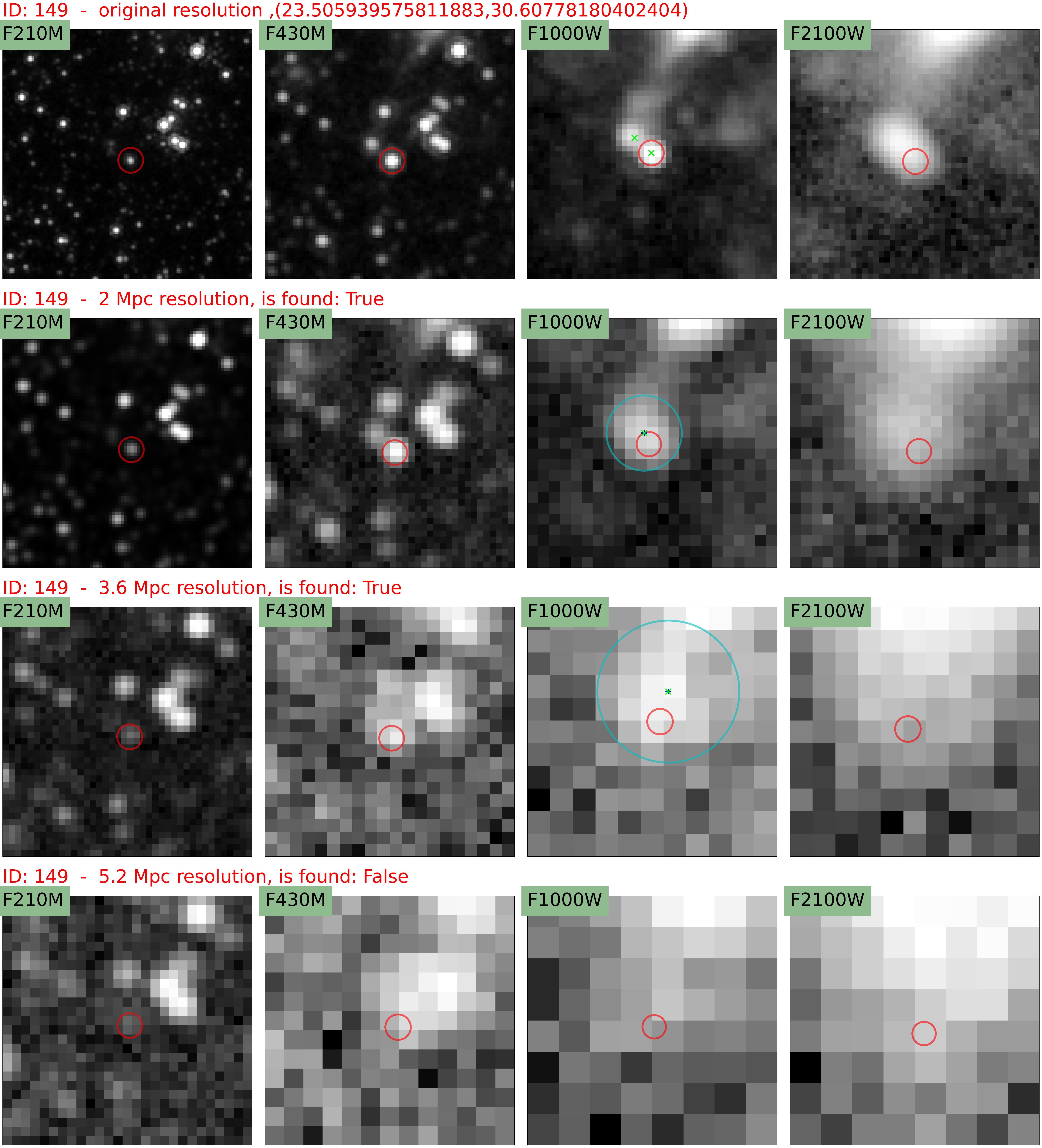}
    \caption{Example of a point-source dominated object in M33 shown at increasing simulated distances, from the native resolution at 0.85~Mpc to 5.2~Mpc. In the original F1000W image two nearby sources are detected (green symbols); the red circle marks the position of object 149, which is the focus of this example. As the images are convolved to larger distances the two sources blend into a single F1000W peak. The peak detected in each convolved image is marked in green, and the cyan circle indicates the search radius used to match detections with the original catalog. At intermediate distances (2–3.6~Mpc) the two sources merge into a single detection, and ID:149 is recovered because it lies closest to the center of the blended peak, while its neighbor is lost due to merging. At 5.2~Mpc both sources are diluted into the diffuse ISM and no distinct 10~$\mu$m peak is detected, resulting in two losses due to dilution. Although source detection in the convolved images is based solely on the 10~$\mu$m data, the corresponding 2, 4, and 21~$\mu$m images are used for selection and classification purposes.}
    \label{fig:merge_dilution}
\end{figure}

Object 149 in Fig.~\ref{fig:merge_dilution} (red circle) provides an example in M33. In the original F1000W image, two nearby dusty sources are detected (green symbols). At intermediate distances (2--3.6~Mpc) the sources merge into a single peak. Object 149 is still recovered because it lies closest to the center of the new blended peak, while its neighbor is lost due to merging. At 5.2~Mpc both sources become diluted into the diffuse ISM and no distinct 10~$\mu$m peak is detected.


In M33 convolved to a 2~Mpc equivalent spatial resolution, most losses are caused by merging (73\%), with the remaining 27\% due to dilution. At 3.6~Mpc merging and dilution contribute roughly equally (43\% and 57\% respectively), while at 5.2~Mpc dilution dominates, accounting for 74\% of the losses. In NGC~300 the small number of lost sources are 
due to merging. In NGC~7793 convolved to 5.2~Mpc, merging accounts for 54.5\% of the losses and dilution for 45.5\%.


\subsection{Impact on CMD-Based Selection}

\begin{figure*}
    \centering
    \includegraphics[width=0.8\linewidth]{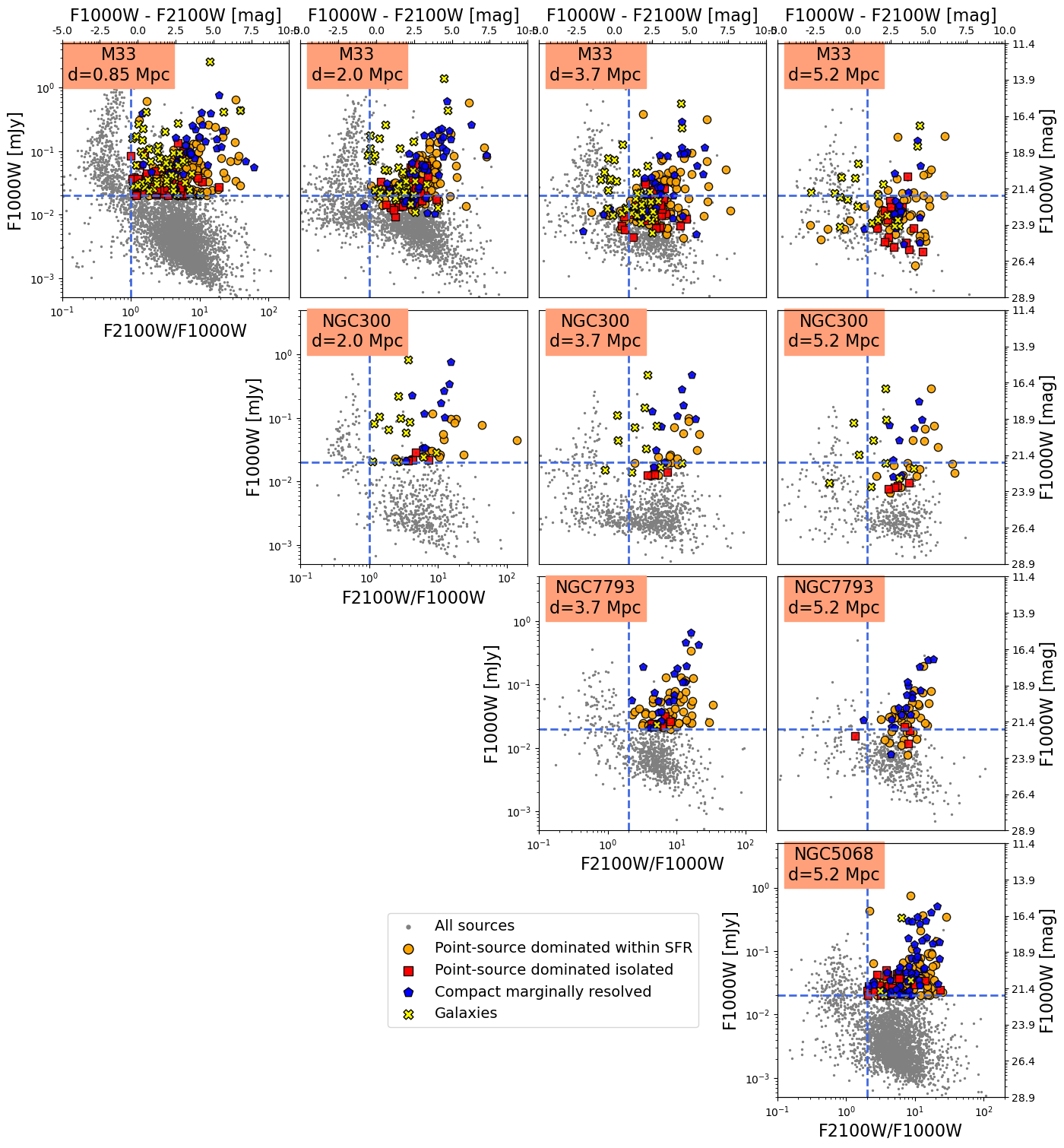}
    \caption{ 
    Set of color–magnitude diagrams showing F2100W/F1000W vs. F1000W in mJy (with AB–magnitude conversions indicated on the top and right axes), for the four galaxies at different resolutions. Each row corresponds to a galaxy (from top to bottom: M~33; NGC~300; NGC~7793 and NGC~5068) and each column corresponds to a specific distance: M~33 at native resolution (0.85 Mpc) and convolved to 2, 3.6 and 5.2 Mpc (top row), NGC~300 at 2, 3.6 and 5.2 Mpc (2nd row), NGC~7793 at 3.6 and 5.2 Mpc (3rd row) and NGC~5068 at 5.2 Mpc (bottom row). Gray points indicate all sources detected at 10~$\mu$m at each resolution, while colored points highlight the dusty young population and background galaxies. 
    The colors of the dusty young population do not change significantly with resolution, although the faintest sources fall below the 10~$\mu$m selection limit at larger distances. 
}
    \label{fig:CMD_categories_convolved}
\end{figure*}

Fig.~\ref{fig:CMD_categories_convolved} shows the F1000W vs. F2100W/F1000W CMD for the three galaxies at both their native (panels along the diagonal) and convolved distances (plots to the right of the diagonal). The colored symbols represent the visually defined categories of dusty young objects defined in Sect.~\ref{subsec:classification}, as well as background galaxies. In the CMDs at convolved distances, the objects successfully matched to sources in the convolved catalogs are marked in color.


A significant fraction of recovered objects fall below the adopted magnitude limit of F1000W $=2\times10^{-2}$~mJy. In M~33, this fraction increases from 24\% at 2~Mpc to 64\% at 3.6~Mpc and 61\% at 5.2~Mpc. Similar trends are seen in NGC~300 and NGC~7793. These sources would not be selected in more distant galaxies because they fall below the magnitude threshold and have large photometric uncertainties.

For sources above the magnitude limit, the mid-infrared colors remain largely stable. Only 
one object in NGC~7793 shifts to the blue side of the CMD and would therefore be excluded from the selection. In this case the color change results from blending with a brighter blue source.
No objects in M33 or NGC~300 shift to the blue side. These results confirm that the colors of dusty young populations are largely insensitive to distance effects.

\begin{figure}
    \centering
      \includegraphics[width=1\columnwidth]{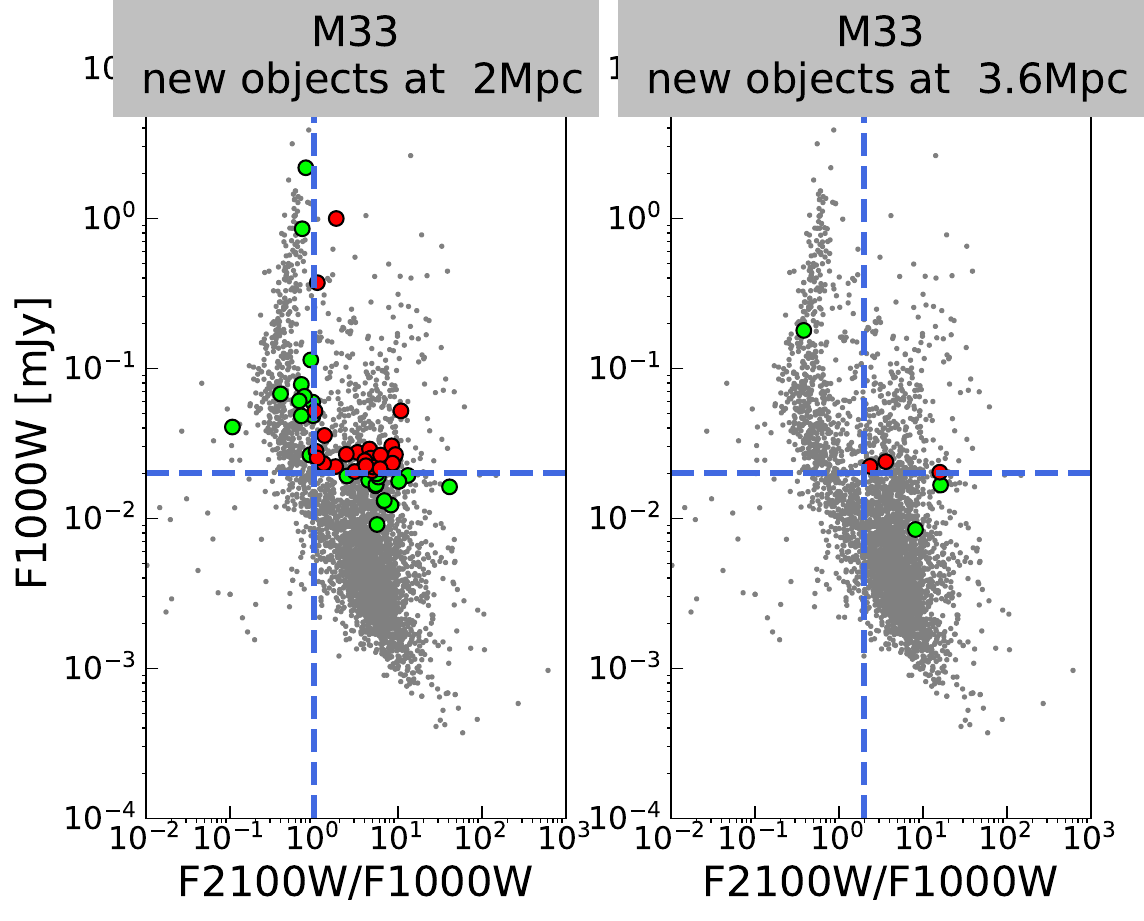}
\caption{Original (green) and final (red) positions in the F1000W vs. F2100W/F1000W CMD of M33 sources that move into the upper-right region of the diagram in the convolved images. Some sources already satisfy the F2100W/F1000W color criterion but are initially too faint to meet the magnitude threshold; after merging with neighboring sources they become brighter and move into the selected region. Others shift from the blue to the red side of the diagram due to small changes in their measured colors.}
\label{fig:new_red}
\end{figure}

Conversely, a modest number of sources move \emph{into} the selected region of the CMD in the convolved images. In M~33 at 2~Mpc, 28 such sources appear (Fig.~\ref{fig:new_red}). Most were originally red but too faint to satisfy the magnitude limit; after merging with nearby sources they become brighter and enter the selected region. A smaller number shift from the blue side of the CMD due to photometric variations. Visual inspection indicates that only one of these 12 sources is likely to be a genuine dusty young object. The remaining sources appear to shift across the color-selection boundary due to small changes in the photometry within the measurement uncertainties. In the mid-infrared, these uncertainties are dominated by the estimation of the complex interstellar background, which can vary slightly as the images are convolved and rebinned. At larger distances the number of newly selected sources decreases substantially. In NGC~300 only two new red sources appear at 5.2~Mpc, while in NGC~7793 sixteen are found.


\begin{table*}
\centering
\caption{Summary of distance effects on the dusty young population: source losses (top) and CMD effects (bottom).}
\label{tab:distance_effects}
\small
\setlength{\tabcolsep}{4pt} 
\renewcommand{\arraystretch}{0.9} 
\begin{tabular}{lccccc}
\hline
\multicolumn{6}{l}{\textbf{(a) Source losses in convolved images}} \\
Galaxy & Conv. dist. & N-objects/N-Original  & Lost (N, \%) & By merging & By dilution\\
\hline
M~33  & 2.0 Mpc & 183/216 &   33 (15\%) & 24 (73\%) & 9 (27\%) \\
M~33   & 3.7 Mpc & 163/216 & 53 (24.5\%) & 23 ($\sim$43\%) & 30 ($\sim$57\%) \\
M~33 & 5.2 Mpc & 117/216 & 99 ( 46\%) & 26 (26\%) & 73 (74\%) \\
NGC~300  & 3.7 Mpc & 30/32  & 2 (6\%) & 2 (100\%) & 0  \\
NGC~300  & 5.2 Mpc & 29/32  & 3 (9\%) & 3 (100\%) & 0   \\
NGC~7793 & 5.2 Mpc & 58/80 & 22 (27.5\%)& 12 (54.5\%) & 10 (45.5\%) \\
\hline
\multicolumn{6}{l}{\textbf{(b) CMD-based effects}} \\
Galaxy  & Conv. dist. & Recovered below limit & Shifted blue (excluded) & New red sources  \\
\hline
M~33   & 2.0 Mpc & 24.5\% & 0 & 28 (16 faint-red, 12 blue → only 1 real) &   \\
M~33   & 3.7 Mpc & 64\% & 0 & 3 (2 red, 1 blue → real)  & \\
M~33 & 5.2 Mp  & 61\% & 0 & 0 &\\
NGC~300  & 3.7 Mpc & 40\% & 0 & 0 &  \\
NGC~300  & 5.2 Mpc & 58.6\% & 0 & 2 &  \\
NGC~7793  & 5.2 Mpc& 24\% &  1 & 16  &\\
\hline
\end{tabular}
\end{table*}

\begin{figure}
    \centering
      \includegraphics[width=0.9\columnwidth]{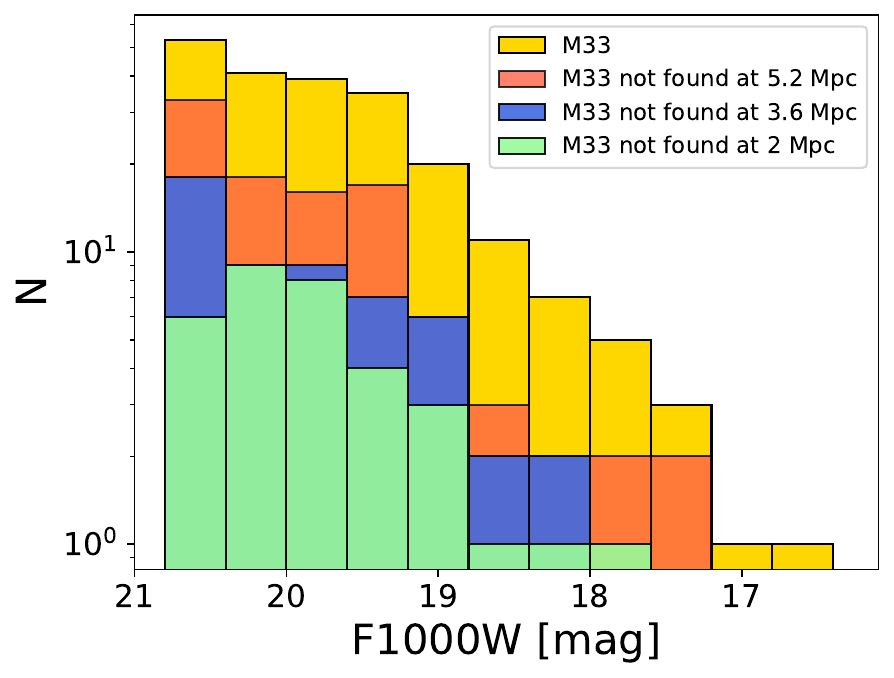}
      \includegraphics[width=0.9\columnwidth]{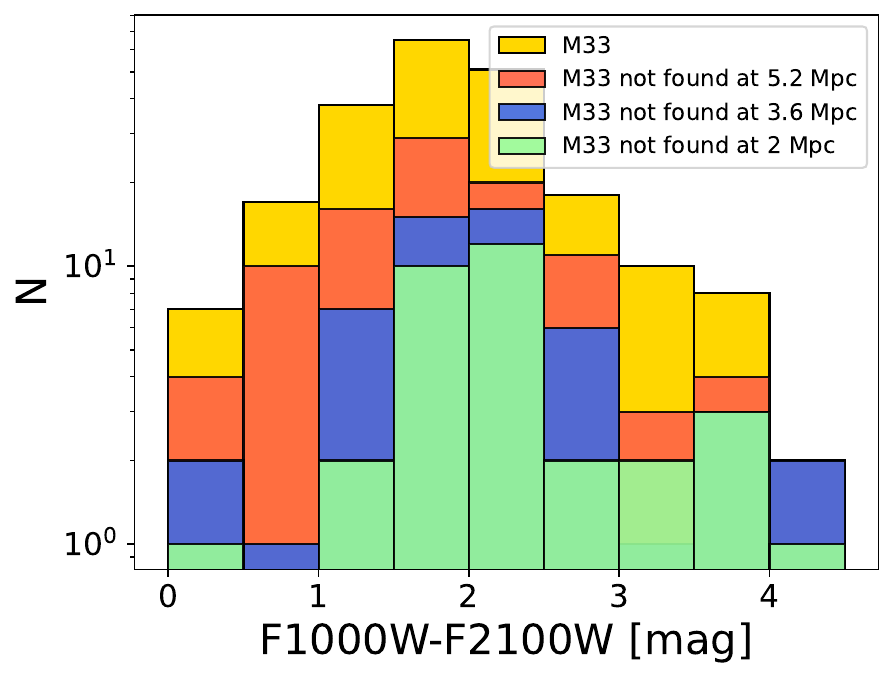}
\caption{
\textit{Top:} Histograms of the F1000W flux for dusty young sources detected in M~33 at native resolution (yellow). The subsets of these sources that are not recovered when the images are convolved to distances of 2, 3.6, and 5.2~Mpc are shown in green, blue, and orange, respectively. The losses occur predominantly among the fainter sources, while the brightest objects remain detectable even at the largest simulated distances. 
\textit{Bottom:} Histograms of the $\mathrm{F1000W}-\mathrm{F2100W}$ color for the same samples, using the same color coding as in the left panel. The similar color distributions indicate that the recovery of sources shows no strong dependence on mid-infrared color.
}
\label{fig:Histograms}    
\end{figure}

Figure~\ref{fig:Histograms} summarizes how distance affects the detection of the dusty young population in M~33. The top panel compares the F1000W distribution of all dusty young sources detected at native resolution with the subsets that are not recovered at 2, 3.6, and 5.2~Mpc. The brightest sources are generally recovered even at the largest simulated distances, indicating that incompleteness primarily affects the fainter end of the population. The bottom panel shows the corresponding $\mathrm{F1000W}-\mathrm{F2100W}$ color distributions, which reveal no significant color dependence in the sources recovered at larger distances. Together, these results indicate that distance primarily reduces completeness through the loss of faint sources, while the color properties of the dusty young population remain largely unaffected.

\subsection{Compact vs Diffuse Dust Emission with Degrading Resolution}
\label{sec:fiffuse_n_compact}

A long-standing question in studies of mid-infrared star-formation tracers is how much of the diffuse emission  actually originates from unresolved star-forming populations rather than from dust heated by an overall interstellar radiation field (older stellar populations). As spatial resolution decreases, compact sources blend together and with surrounding diffuse emission, potentially altering the fraction of the total mid-infrared luminosity that can be directly associated with young stellar populations.

To quantify this effect, we examine how the fraction of F2100W emission attributed to compact dusty star-forming objects changes as spatial resolution degrades. We use the M33 dataset at its native resolution and the versions of the images convolved to simulate observations at distances of 2 and 3.6~Mpc. At each distance, the F2100W image is first smoothed to a physical resolution of 100~pc. Regions of similar surface brightness are then identified, and the galaxy is subdivided into between 3 and 30 spatial patches using KMeans clustering, ensuring that each patch covers at least five resolution elements.

Within each patch we measure two quantities: (1) the total F2100W flux density contained in the smoothed image, and (2) the summed F2100W flux density of all compact sources identified in our catalogs. Compact star-forming sources are selected using the same color criterion applied throughout this paper, $\mathrm{F}_{\nu}(21\,\mu{\rm m})/\mathrm{F}_{\nu}(10\,\mu{\rm m})>1$, but without imposing a flux limit so that faint red sources are retained. As discussed in Sects.~\ref{sect:other_catalogs} and \ref{sect:other_catalogs_5068}, these objects are likely associated with young dusty star formation.

\begin{figure}
    \centering
    \includegraphics[width=0.9\columnwidth]{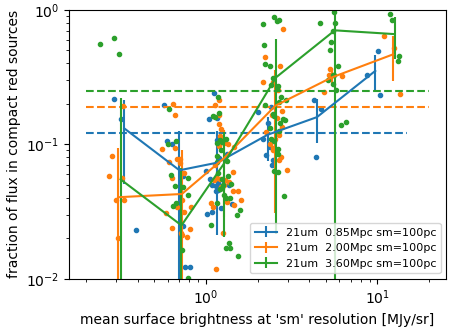}
    \caption{Fraction of F2100W emission which is identified as compact star-forming sources, as a function of 100$\;$pc surface brightness, and at 3 different distances.  The fraction of emission in compact sources is higher in brighter regions of the image, and increases as the distance increases and resolution degrades.  See text for details of the calculation -- the dashed lines show the fraction over the entire image.  The solid lines show the median fraction in selected patches of the galaxy, i.e. for all 100pc-sized patches within a certain surface brightness range, the mean and standard deviation of flux in compact sources among those patches is calculated.  The solid vertical error bars indicate each standard deviation, and the solid lines join the means.  }
    \label{fig:fraction_flux_YSOs}
\end{figure}

Figure~\ref{fig:fraction_flux_YSOs} shows the fraction of F2100W emission attributed to compact star-forming sources as a function of the average 100~pc-scale surface brightness in each patch. Individual patches are shown as points, while the median and RMS within each surface-brightness bin are indicated by solid curves. At a given surface brightness, the fraction of emission contained in compact sources exhibits significant scatter.

Despite this scatter, two systematic trends are evident. First, regions with higher surface brightness—typically corresponding to spiral arms and active star-forming complexes contain a larger fraction of their mid-infrared emission in compact sources. Second, as the spatial resolution decreases, a larger fraction of the total emission is attributed to compact sources. This behavior likely reflects the strong clustering of star formation: as the resolution degrades, nearby compact sources merge into more luminous objects that are still identified as compact star-forming sources at the lower resolution.

Interestingly, the opposite trend appears in regions of lower surface brightness. In these environments, which correspond to more quiescent areas of the galaxy, the fraction of emission attributed to compact sources decreases as the resolution degrades. This suggests that in these regions the emission becomes increasingly dominated by diffuse dust rather than by unresolved compact star-forming sources.

These results indicate that the apparent balance between compact and diffuse mid-infrared emission depends strongly on both spatial resolution and the local star-formation environment. In regions of active star formation, much of the emission attributed to diffuse dust at coarse resolution may in fact arise from unresolved clusters of compact star-forming sources. In contrast, in lower surface-brightness regions a larger fraction of the emission likely originates from dust heated by more widely distributed stellar populations.

\section{Luminosity Functions and Approximate Mass Scales}

\subsection{10$\mu$m Luminosity Function}
\label{sec:LF}
\begin{figure*}
      \includegraphics[width=1\linewidth]{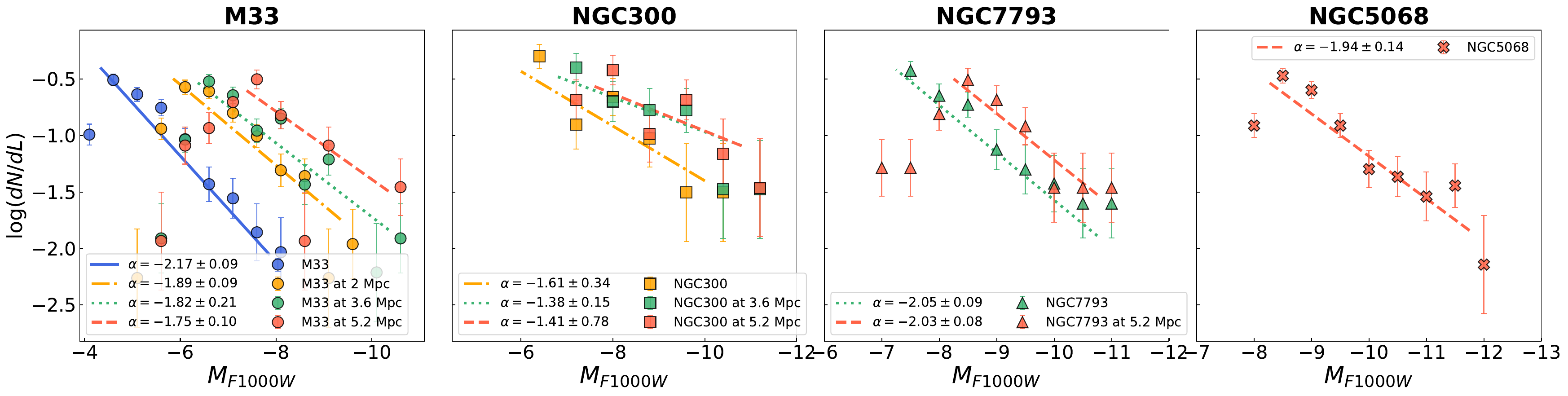}
    \caption{ 
F1000W luminosity functions of the dusty young population in M~33, NGC~300, NGC~7793, and NGC~5068 (left to right) at the native and convolved resolutions. As the images are degraded, the LF shifts toward brighter magnitudes due to the blending of multiple sources, while the number of faint sources decreases. The slopes of the bright end remain largely unchanged with distance. M~33, NGC~7793, and NGC~5068 exhibit LF slopes of $\alpha \sim 2$, whereas NGC~300 shows a flatter LF with $\alpha \sim 1.6$. The bin uncertainties are estimated assuming Poisson counting statistics. The uncertainties in the fitted LF slopes are derived from the standard error of the linear regression. 
}
    \label{fig:LF}    
\end{figure*}

Figure~\ref{fig:LF} shows the F1000W luminosity functions (LFs) of the dusty young population in M~33, NGC~300, NGC~7793, and NGC~5068, computed at both the native and convolved distances. As the images are degraded, the LF shifts toward brighter magnitudes due to the blending of multiple sources at lower spatial resolution. At the same time, the number of objects in the faintest bins decreases, as low-luminosity sources become increasingly difficult to distinguish from the diffuse ISM background in these star-forming regions.

The LF is well described by a power-law function of the form $dN/dL \propto L^{-\alpha}$ \citep[e.g.,][]{whitmore99, fall06}. Following this parameterization, we fit a linear relation to the LF from the peak of the distribution toward brighter magnitudes to estimate the slope $\alpha$. M~33, NGC~7793, and NGC~5068 exhibit LF slopes of $\alpha \sim 2$, with only small variations as the images are convolved to larger distances ($\Delta \alpha < 0.4$). This behavior is consistent with a scale-free luminosity distribution similar to that observed for young star clusters in nearby galaxies \citep[e.g.,][]{whitmore14a, krumholz19, Rodriguez25}.

In contrast, NGC~300 shows a flatter LF with $\alpha \sim 1.6$, indicating a relatively larger fraction of luminous sources compared to the other galaxies. This difference could reflect intrinsic variations in the dusty young population in the limited area of the galaxy sampled or observational effects such as blending or incompleteness at the faint end.

\subsection{Approximate mass scales}
\label{sec:masses}

Estimating stellar masses for dusty sources is not straightforward. The objects identified here span a range from individual MYSOs to small embedded groups and compact young clusters, and their observed mid-infrared emission depends not only on stellar mass, but also on extinction, viewing geometry, circumstellar dust, multiplicity, and blending. A robust treatment would require SED fitting and radiative-transfer modeling, ideally informed by source morphology. Such an analysis is beyond the scope of the present paper, but will be an important focus of future work using the new Cycle~5 JWST 9-band NIRCam and MIRI imaging survey of M33 (PI Rosolowsky), spanning 1--21$\mu$m.

Here, we take an initial step by deriving approximate bounds on the characteristic masses spanned by the sample using two simple empirical approaches. First to estimate the lower bound of masses, we compute the 24$\mu$m mass-to-light ratios based on MYSOs in the \textit{Spitzer} SAGE-LMC survey whose masses were estimated through radiative transfer SED modeling \citep[e.g.,][]{whitney08,gruendl09,seale14}, assuming each source is dominated by a single star:
$\rm Log(M)= 0.29~log (L_{24\mu m}) +0.63 $
This provides an estimate of the mass of the dominant embedded object that would be required to produce the observed mid-infrared emission. The SAGE 24$\mu$m flux density is reasonably correlated with the SAGE MYSO SED-fitted mass (Spearman correlation coefficient of 0.83), and F2100W is very well correlated to MIPS 24$\mu$m for this population as discussed in Appendix~\ref{append:color_corr}, so the F2100W flux should serve as a reasonable proxy for the 24$\mu$m emission used in the original SAGE MYSO calibration.  

Second to estimate a likely upper bound of masses, we assume that the near-infrared emission traces the stellar continuum of a compact young cluster. Here, we apply a characteristic mass-to-light ratio at $2\mu$m based on the PHANGS-HST star cluster catalogs with ages $<$ 3~Myr, following \citet{Rodriguez25} but only for NGC~5068:
$ \rm M_{F2000W}=-1.81~log (M/M_\odot) +1.54$,
which assumes that sources have relatively well-sampled stellar mass functions (i.e., total masses $\gtrsim$10$^4$M$_\odot\,$). This conversion provides an approximate estimate of the total stellar mass of the system in the regime where the observed source corresponds to a full population of stars formed instantaneously. 

These two approaches should roughly bracket the plausible mass scale of the sources, ranging from the mass of a dominant embedded massive star to the total stellar mass of a compact embedded stellar system. The two estimators intentionally probe different physical components and spatial scales: the 21$\mu$m emission traces warm circumstellar and interstellar dust on larger physical scales, whereas the 2$\mu$m emission isolates more compact stellar continuum sources. Consequently, the two estimates are not expected to agree. 

\begin{figure*}
 \includegraphics[width=1\linewidth]{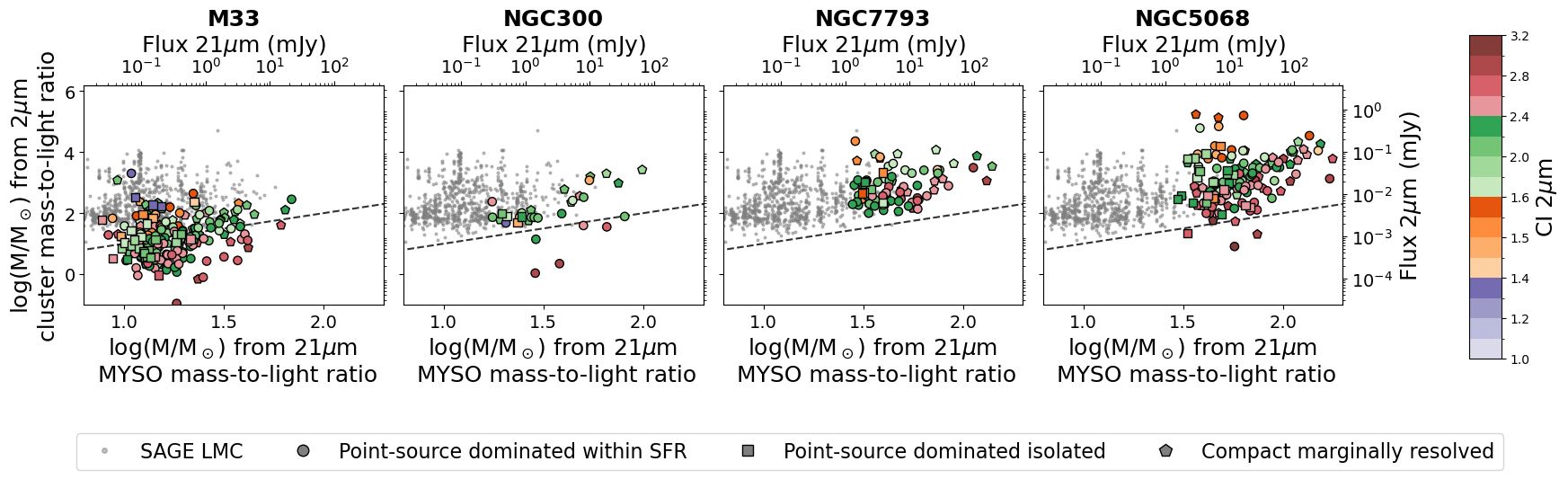}
    \caption{Constraints on the mass range spanned by the sample of young dusty compact sources presented here using two limiting approaches. Masses shown on the x-axis are derived from SAGE MYSOs based on the 21$\mu$m luminosity, 
    while masses on the y-axis are computed using the mass-to-light ratio of young ($<$ 3~Myr) PHANGS–HST star clusters at 2$\mu$m.  Different symbols correspond to different visual classifications (Section 4.3) and are color coded based on the F200W concentration index. For reference, LMC MYSOs from SAGE are shown where the x-axis values correspond masses from radiation transfer modeling of the SED. The gray dashed line correspond to to the one-to-one relation.} 
    \label{fig:masses}   
\end{figure*}

In Fig.~\ref{fig:masses} we compare the masses estimated using the two methods. For reference, we also overplot the SAGE LMC MYSOs in gray, where the masses shown on the x-axis are derived from radiative transfer modeling of the SED, while the masses on the y-axis are estimated using the 2$\mu$m mass-to-light ratio for stellar clusters.

The M33 sources overlap the regime occupied by the SAGE sample, providing reassurance that our selection is recovering sources with properties broadly consistent with known individual MYSOs. The fluxes of the sources gradually increase with galaxy distance, as expected from the increasing luminosity threshold imposed by the fixed flux limits and from greater source blending at coarser physical resolution. 

For M33 and NGC300, the overlap with the SAGE MYSO sources suggests that the 21 $\mu$m/SAGE-MYSO estimator provides the more appropriate mass scale (10-70 and 20-99 M$_{\odot}$, respectively). The 2~$\mu$m/cluster estimator can still be considered as an upper-bound or total-system estimate, but it should not be interpreted literally particularly for these low-mass, sparsely populated systems.

For NGC7793 and NGC5068, the sources move increasingly away from the SAGE MYSO regime, and the 21 $\mu$m/SAGE-MYSO estimator becomes less reliable as a physical mass estimate. In these more distant galaxies, the individual MYSOs are more likely to be blended. The 2 $\mu$m/cluster estimator is therefore more relevant as an approximate total stellar-system mass (100-2$\times$ 10$^{4}$ and 10-2 $\times$ 10$^{5}$ M$_{\odot}$, respectively), but still only as an order-of-magnitude guide due to factors such as dust attenuation.

The M33 sources are offset toward lower 2$\mu$m fluxes compared to the SAGE sample, but this offset is largely driven by differences in the underlying imaging. The SAGE values are based on 2MASS $K_s$ photometry at 2.16$\mu$m with a PSF FWHM of $\sim$2--3\arcsec, corresponding to physical scales of $\sim$0.5--0.7~pc at the distance of the LMC, whereas the JWST F200W PSF probes $\sim$0.29~pc at the distance of M33. The factor of two higher spatial resolution of JWST isolates compact structures on much smaller physical scales and reduces contamination from neighboring sources and diffuse extended emission, leading to systematically lower measured near-infrared fluxes for otherwise comparable objects. Although there is also an approximately factor of two difference in physical resolution in the opposite direction between the \textit{Spitzer} 24$\mu$m observations of the LMC (6\arcsec, corresponding to $\sim$1.5~pc) and the JWST 21$\mu$m observations of M33 (0\farcs67, corresponding to $\sim$2.8~pc), the impact on the inferred masses is smaller because the SAGE-based MYSO calibration scales approximately as $M\propto L^{1/3}$, whereas the cluster-based masses scale linearly with luminosity ($M\propto L$). As a result, differences in spatial resolution and source blending at 2$\mu$m are more obvious in Fig.~\ref{fig:masses}.

\section{Discussion}
\label{sec:discussion}

\subsection{Comparison with Literature Catalogs in M~33}
\label{sect:other_catalogs}
A key goal of this work is to understand the selection of dusty star forming objects that span the range from single intermediate-mass YSOs through massive star clusters. 

Star clusters in M33 have been previously identified by \citet{sarajedini07} and re-analyzed including infrared data by \citet{moeller22}.  In the region imaged here, there are 9 of those clusters with ages $<$10$^7$yr.  

\citet{ren21} used B through K ground-based photometry to identify candidate red supergiant stars. 87 of these are also found in our F1000W-based catalog,
but \textbf{none} are included in our dusty young object sample.

We also compared our sample of dusty young objects in M33 with the PHATTER cluster catalog 
\citep{2022ApJ...938...81J}, which identified clusters using HST imaging. Of the 161 PHATTER clusters located within our observed field of view, only two are found in close projected proximity to our dusty young sources. 
Note that sources with nebular emission were excluded from the PHATTER cluster catalog, almost certainly excluding many of our sources which are located in complex star-formation regions with significant diffuse emission.

\begin{figure}
    \centering
    \includegraphics[width=1\linewidth]{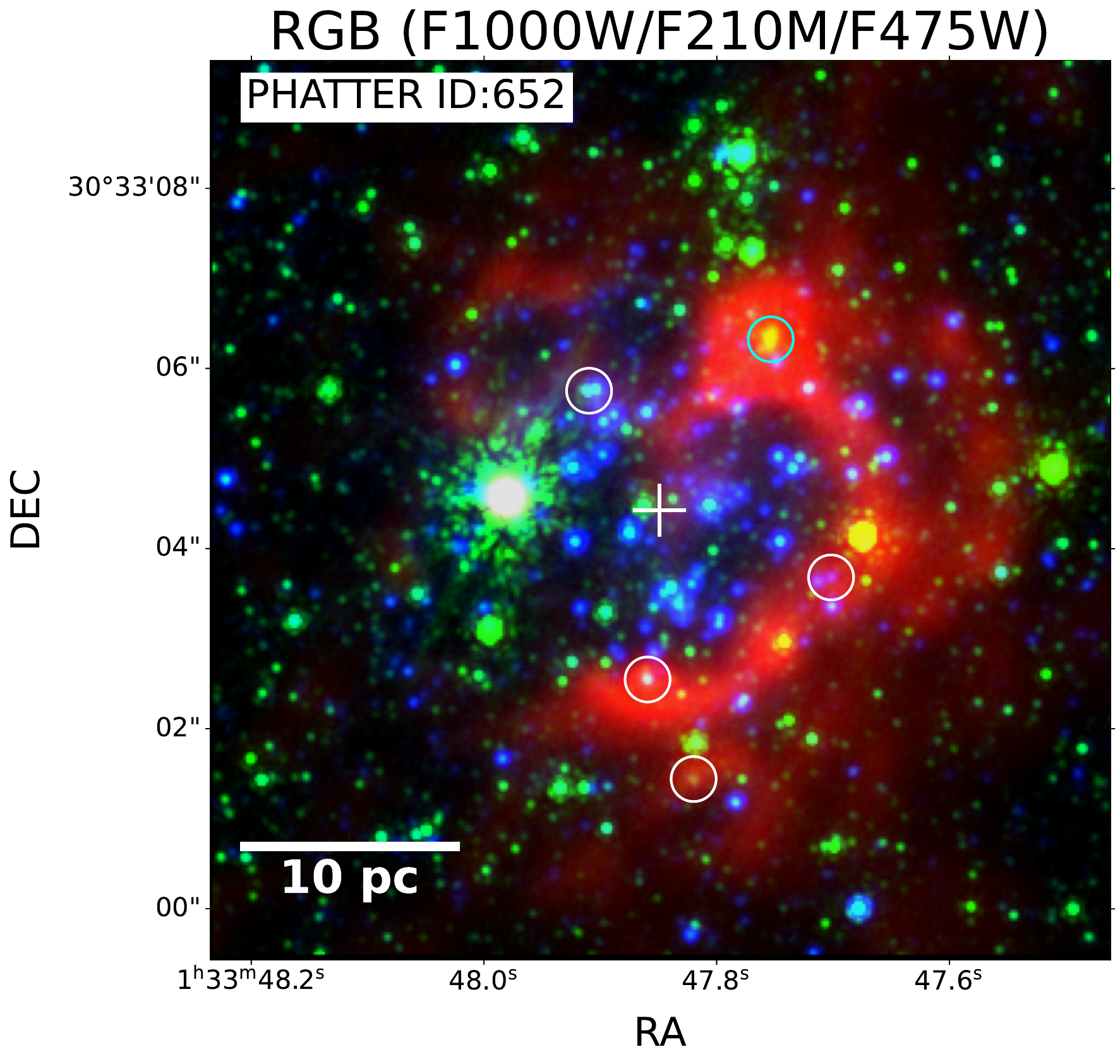}
    \caption{color image of the PHATTER's cluster ID: 652 (the white cross shows the cluster's center). Blue: HST F475W, green: JWST F210M, red: JWST F1000W. White circles are objects identified as point-source dominated and the cyan is an objects classified as compact marginally resolved in the current work.}
    \label{fig:Phatter_cluster}
\end{figure}

Interestingly, we found that PHATTER cluster ID~652 (Fig.~\ref{fig:Phatter_cluster}) is spatially associated with a group of four point-source dominated sources and one bright compact marginally resolved source identified in our catalog.
In this region, the PHATTER cluster appears to be triggering star formation along an arc-like structure where all of our dusty young objects are located. The PHATTER cluster has an estimated age of $\sim$4.5~Myr \citep{2022ApJ...928...15W}; this cluster has likely already expelled its natal gas through stellar feedback, shaping the local ISM and plausibly compressing the surrounding material \citep[see e.g.][]{rahner}. This process may have contributed to the formation of the observed arc-like structure and to the subsequent triggering of star formation along it.

The very low overlap between the PHATTER cluster catalog and our dusty objects is expected, since PHATTER clusters are typically older than $\sim$4~Myr and sources with nebular emission were excluded \citep{2022ApJ...928...15W}, whereas the objects targeted in this work are expected to be significantly younger ($\lesssim$1~Myr). The two catalogs also exhibit markedly different spatial distributions: very few PHATTER clusters are located within the ISM structures traced by the 10~$\mu$m emission, where our dusty objects are preferentially found.


We also compare our catalog with the YSO candidates identified by \citet{Peltonen2024} using JWST/MIRI 5.6 and 21~$\mu$m imaging. We adopt the published YSO positions and perform aperture photometry following the same procedure described in Sect.~\ref{sec:photometry}, in order to place these sources in the 10~$\mu$m versus 21~$\mu$m/10~$\mu$m CMD (Fig.~\ref{fig:CMD-NGC5068_catalogs}, left panel).
Of the 465 YSO candidates from \citet{Peltonen2024} that fall within our 10~$\mu$m field of view, only 179 lie within our upper-right CMD selection region. Among these, we recover 89 objects in our catalog. The majority of the remaining sources lie within very bright and crowded 10~$\mu$m regions, where multiple 10~$\mu$m sources are blended together, preventing us from detecting individual 10~$\mu$m peaks that may correspond to objects in the \citet{Peltonen2024} catalog.
Most of the YSO candidates that fall below our 10~$\mu$m detection threshold (280 sources) do not show a clear associated 10~$\mu$m counterpart and are likely detected primarily at other wavelengths. Conversely, we identify a substantial number ($>$100) of dusty young objects in our catalog that are not present in the \citet{Peltonen2024} sample. Despite the discrepancies between the two catalogs, Fig.~\ref{fig:CMD-NGC5068_catalogs} shows that the 21~$\mu$m/10~$\mu$m colors of the YSO candidates from \citet{Peltonen2024} are consistent with the color criteria adopted in this work.

\subsection{Comparison with PHANGS's catalogs in NGC~5068}
\label{sect:other_catalogs_5068}

\begin{figure*}
    \centering
    \includegraphics[width=1\linewidth]{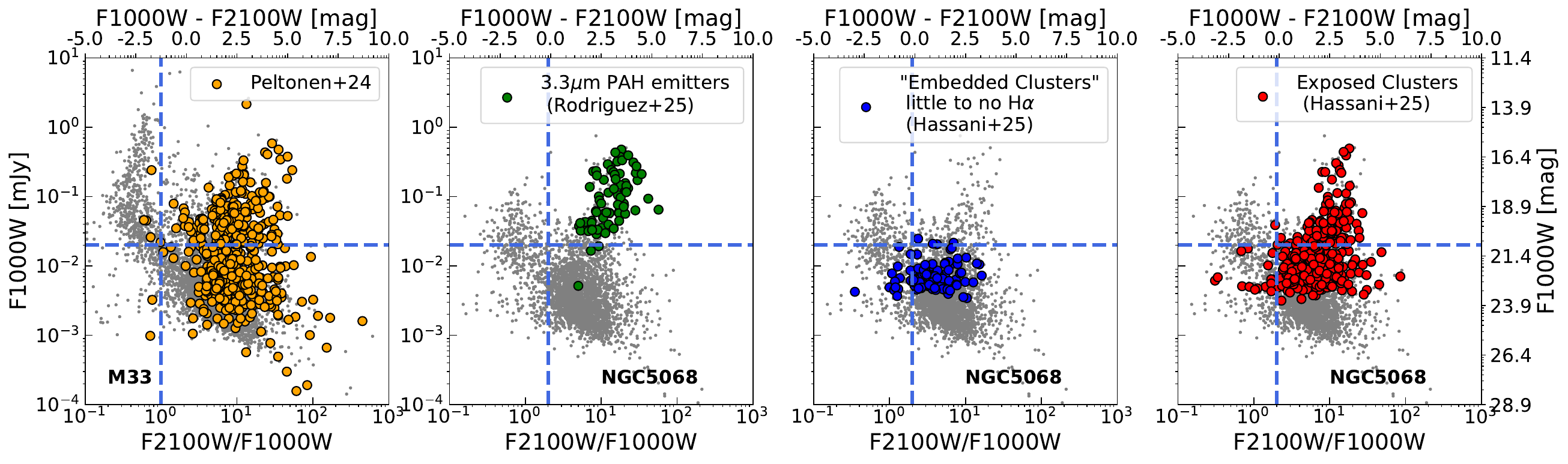}
    \caption{CMDs for M~33 (left) and NGC~5068 (second, third, and fourth panels), showing different catalogs of dusty young objects compiled from the literature, used for comparison in this work.}
    \label{fig:CMD-NGC5068_catalogs}
\end{figure*}

The dusty-young population of NGC~5068 has been studied in previous works. In particular \cite{Rodriguez25} detected 77 dust-embedded star clusters presenting 3.3$\mu$m PAH emission. 
We took the positions of that catalog and estimated the photometry fluxes in F1000W and F2100W using exactly the same procedure we are using in this work, which is explained in Sect.~\ref{sec:photometry}. Then we placed them in the 10~$\mu$m vs. 21~$\mu$m/10~$\mu$m CMD (Fig.~\ref{fig:CMD-NGC5068_catalogs}, second panel) in order to check if they fall in the selection region (top right), and therefore pass our initial selection cut. From the 77 PAH emitters we found 74 in this region, and 60 of them match in sky positions with objects selected in this work. As the PAH emitters were selected using the F335M NIRCam image which has a better resolution (0\farcs1) than our selection band (F1000W, 0\farcs328), some of our 10~$\mu$m objects correlate with more than one 3.3~$\mu$m PAH emitter. This shows a good correlation between both methods to detect dusty young objects.  As in the case of the YSOs candidates from \cite{Peltonen2024} mentioned in Sect~\ref{sect:other_catalogs}, most of the PAH emitters for which we did not find a match in the 10$\mu$m dusty objects list are located in bright and crowded 10$\mu$m regions, affected by blending, and due to this we were not able to detect a 10~$\mu$m peak corresponding to a a 3.3~$\mu$m source, although the region is  bright at 10~$\mu$m.

We performed the same procedure using the cluster catalogs from \cite{Hassani25}, where they put young clusters in two different categories. The Embedded Clusters (Fig.~\ref{fig:CMD-NGC5068_catalogs}, third panel), defined as mid-IR objects which present a strong 3.3$\mu$m PAH feature, and show little to no HST-H$\alpha$ emission.
On the other hand the Exposed Clusters from \cite{Hassani25} (Fig.~\ref{fig:CMD-NGC5068_catalogs}, right panel) are young star clusters detected in both mid-IR and H$\alpha$ emission.

In Fig.~\ref{fig:CMD-NGC5068_catalogs} (third panel), we note that although most of the embedded clusters from \cite{Hassani25} exhibit red F2100W/F1000W colors, consistent with our color criteria, only four of them are bright enough at 10~$\mu$m to be included in our selection. All four objects are recovered in our catalog.
In the right panel of Fig.~\ref{fig:CMD-NGC5068_catalogs}, we show the exposed clusters from \cite{Hassani25}. This catalog contains 379 objects, of which 97 fall within our selection region (i.e., satisfy our initial selection criteria of F2100W/F1000W $\geq$ 2 and $\rm F1000W \geq 2 \times 10^{-2}$~mJy). We find that 100 out of the 379 objects have positional matches with sources in our dusty-object catalog. The small excess of three objects relative to the 97 sources within the selection region is due to slight differences in the source coordinates between the two catalogs. These coordinate offsets lead to small differences in the aperture placement used for photometry, which can shift some objects marginally above or below the selection boundaries.

\section{Summary and Conclusions} 
\label{sec:conclusion}

Based on JWST 2, 4, 10, and 21~$\mu$m observations, we selected a population of dusty young sources in four nearby galaxies spanning distances of $\sim$1–5~Mpc. This sample allows us to assess how increasing distance and decreasing spatial resolution affect the identification and characterization of dusty young stellar populations, including massive young stellar objects (YSOs). Our main findings can be summarized as follows:

\begin{itemize}


    \item Using the SAGE-LMC MYSO catalog as a benchmark, we simulated the effects of increasing distance and source blending in JWST observations to guide the selection of compact dusty young sources in nearby galaxies. These experiments demonstrate that mid-infrared color selection remains robust out to at least 5 Mpc and suggest that systems whose infrared emission is dominated by a single MYSO can typically be identified and characterized out to distances of $\sim$3 Mpc, beyond which source blending becomes increasingly important (Sect.~\ref{sec:SAGE_experiments}).

    \item The most effective strategy for identifying dusty young sources in galaxies out to distances of 5~Mpc combines mid-IR, near-IR, and morphological information. The $\rm F1000W$ versus $\rm F1000W-F2100W$ CMD efficiently separates the dusty young population from evolved stellar populations, while near-IR colors (e.g., $\rm F200W-F444W$) and the 2$\mu$m concentration index provide additional constraints for removing contaminants such as massive main-sequence stars. In particular, dusty young sources typically exhibit $\rm F1000W-F2100W \gtrsim 0.5$,
    red near-IR colors $\rm F200W-F444W \gtrsim 1$, and concentration indices $\rm CI \gtrsim 1.4$. Contaminants are primarily background galaxies, which can generally be identified and removed through visual inspection (Sec.~\ref{sec:classification}--\ref{sec:final_sample}). 

    \item Applying these selection criteria, we identify a total of 467 dusty young sources across our sample of four galaxies out to distances of 5~Mpc, including 216, 32, 80, and 139 sources in M~33, NGC~300, NGC~7793, and NGC~5068, respectively. These objects comprise two main classes: point-source dominated sources, whose IR emission may be dominated by a single massive YSO, and compact marginally resolved sources, which may host multiple massive YSOs and young stars (Sect.~\ref{sec:final_sample}).

   \item Overall, we find that the intrinsic colors of the dusty young population remain largely stable with increasing distance, indicating that the selection is robust against resolution effects. The primary impact of distance is instead a progressive loss of sources due to blending and dilution into the diffuse ISM, which becomes significant at the largest distances and leads to the loss of up to $\sim$50\% of the sample at 5.2~Mpc. A smaller secondary effect is the introduction of a modest level of contamination from intrinsically blue sources that appear red due to blending or photometric uncertainties, although this remains limited ($\sim$5\% in M33 at 2 Mpc) and becomes negligible in the more distant systems. Despite these effects, the brightest dusty young sources (F1000W $<$ 19 mag) are generally recovered even when the data are degraded to simulate larger distances (Sect.~\ref{sec:dusty_pop_at_convolved_images}).

    



    \item The luminosity function of the dusty young population remains largely unchanged with increasing distance, with a slope of $\sim$–2, although it shifts toward brighter magnitudes due to source blending and the loss of fainter objects (Sect.~\ref{sec:LF}).

    \item Using two complementary empirical mass estimators, we bracket the characteristic mass scale of the dusty young population between the mass of a dominant embedded massive star and the total stellar mass of a compact young stellar system. A 21~$\mu$m-based calibration derived from SAGE-LMC MYSOs provides a lower-bound estimate, while a 2~$\mu$m cluster mass-to-light ratio derived from PHANGS-HST young clusters provides an approximate upper-bound estimate. For the nearest galaxies (M33 and NGC~300), the sources overlap the SAGE MYSO regime and are consistent with masses of $\sim$10--70 and 20--100~M$\odot$, respectively. In the more distant galaxies (NGC~7793 and NGC~5068), where source blending becomes increasingly important, the inferred total stellar-system masses span $\sim$100--2$\times10^{4}$ and 10--2$\times10^{5}$~M$\odot$, respectively, indicating that the sample bridges the regime between individual massive embedded stars and compact young stellar systems (Sect.~\ref{sec:masses}).

    

    \item As spatial resolution degrades with distance, an increasing fraction of the total mid-IR 21$\mu$m flux is attributed to compact dusty sources, as intrinsically clustered star-forming regions blend into fewer, more luminous objects that remain classified as compact. In contrast, in less active regions, compact sources tend to merge into the diffuse emission, leading to a decreasing fraction of flux associated with compact sources as resolution decreases (Sect.~\ref{sec:fiffuse_n_compact}).

    \item Our 10~$\mu$m–based selection recovers most ($\sim$80\%) of the 3.3~$\mu$m PAH emitters in NGC~5068 identified by \citet{Rodriguez25}, with the few missing sources largely explained by blending in crowded 10~$\mu$m regions. This strong overlap provides independent validation that our selection criteria efficiently identify dust-embedded young stellar sources (Sect.~\ref{sect:other_catalogs_5068}).

\end{itemize}

Taken together, these results provide a practical framework for identifying and characterizing dusty young stellar populations with JWST in nearby galaxies out to distances of at least 5 Mpc. By spanning a broad range of physical scales--from systems potentially dominated by individual massive embedded stars to compact young stellar systems--this work bridges the gap between Galactic and Local Group studies of massive star formation and investigations of dust-embedded star formation in more distant galaxies.

\appendix

\section{Examples of selected sources}
\label{Ap:fig_examples}
\begin{figure*}
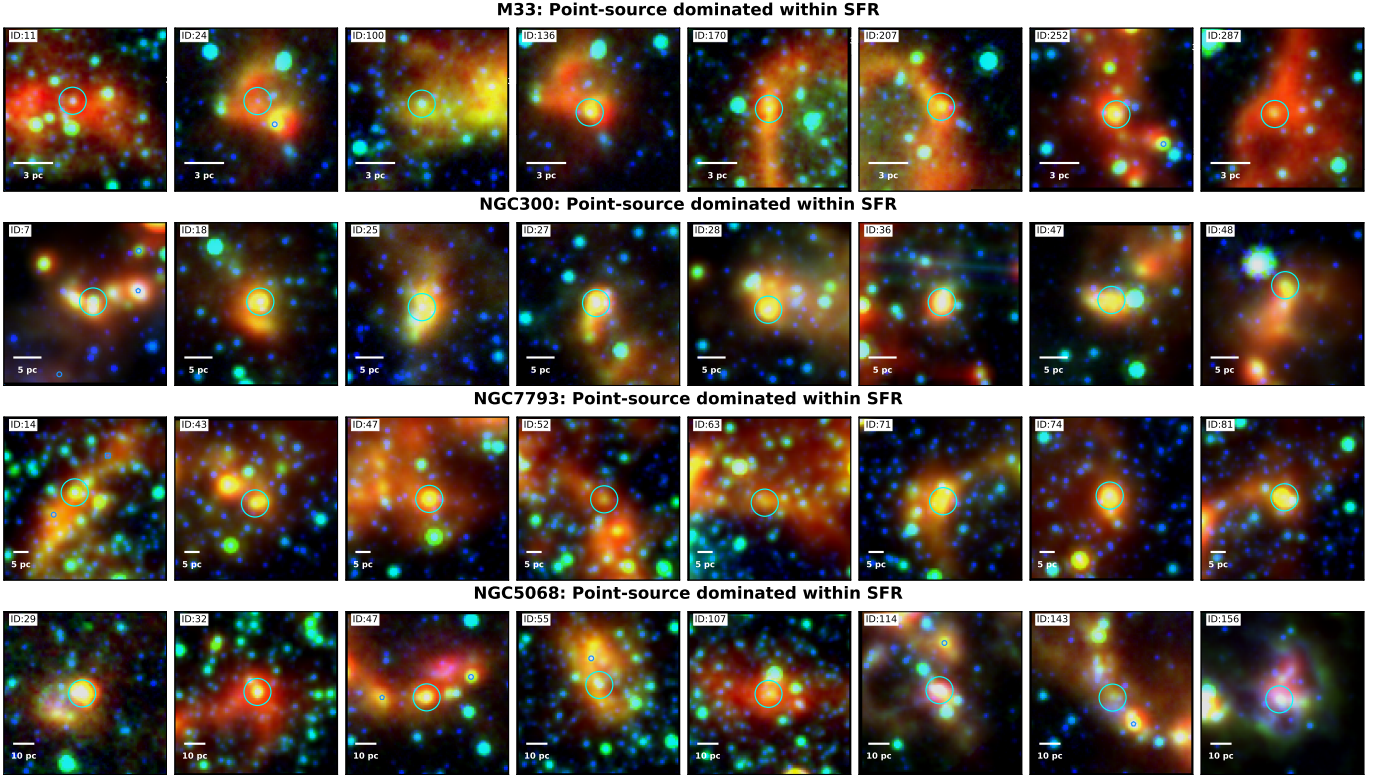

    \centering
    \largefig{\includegraphics[width=1\linewidth]{categories_point_source_SFR_RGB_10_4_2_m33_examples.pdf}\\
    \includegraphics[width=1\linewidth]{categories_point_source_SFR_RGB_10_4_2_ngc0300_examples.pdf}\\
    \includegraphics[width=1\linewidth]{categories_point_source_SFR_RGB_10_4_2_ngc7793_examples.pdf}\\
    \includegraphics[width=1\linewidth]{categories_point_source_SFR_RGB_10_4_2_ngc5068_exaamples.pdf}}
    \caption{Examples of point-source dominated sources within star-forming regions.
    Blue: F210M (M~33) or F200W (NGC~300,NGC~7793 and NGC~7793) / Green:F430M (M~33), F444W (NGC~300 and NGC~7793) or F360M (NGC~5068)/, Red: F1000W. Cutout are 3$\times$3 arcsec$^{2}$ which is equivalent to $\sim$13.6$\times$ 13.6 pc$^{2}$ at the distance of M~33, $\sim$30$\times$ 30 pc$^{2}$ (NGC~300), $\sim$52$\times$ 52 pc$^{2}$ (NGC~7793) and  $\sim$75.5$\times$ 75.5 pc$^{2}$ (NGC~5068).
    The large central circles indicate the objects which ID correspond to the number in the upper left corner, while other objects of the dusty young population in the same field are shown by smaller symbols: circles for point-source dominated sources within SFRs, squares for point-source dominated isolated sources, and pentagons for compact marginally resolved sources. }
    \label{fig:YSO_SFR}
\end{figure*}

\begin{figure*}
    \centering
    \largefig{\includegraphics[width=1\linewidth]{categories_point_source_Isolated_RGB_10_4_2_m33_examples.pdf}\\
    \includegraphics[width=0.5\linewidth]{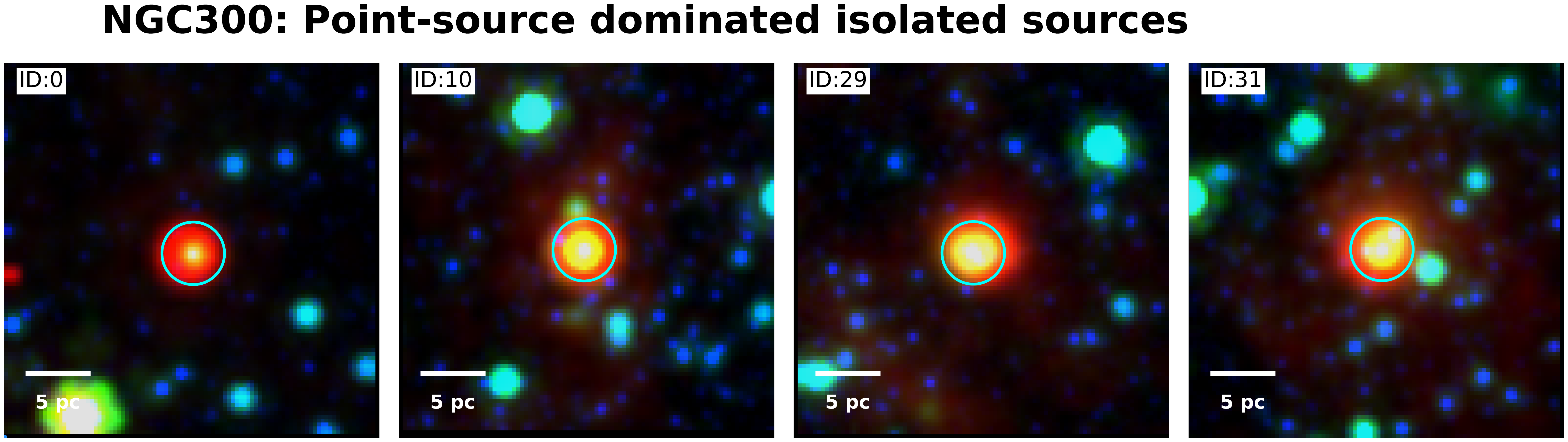}\\
    \includegraphics[width=0.8\linewidth]{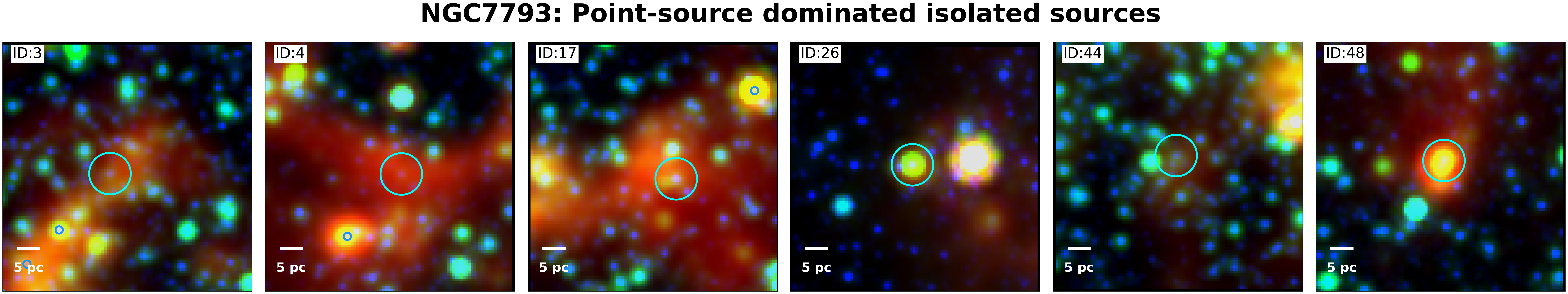}\\ \includegraphics[width=1\linewidth]{categories_point_source_isolated_RGB_10_4_2_ngc5068_examples.pdf}}
    \caption{Examples of point-source dominated isolated sources.
    Blue: F210M (M~33) or F200W (NGC~300,NGC~7793 and NGC~7793) / Green:F430M (M~33), F444W (NGC~300 and NGC~7793) or F360M (NGC~5068)/, Red: F1000W. Cutout are 3$\times$3 arcsec$^{2}$ which is equivalent to $\sim$13.6$\times$ 13.6 pc$^{2}$ at the distance of M~33, $\sim$30$\times$ 30 pc$^{2}$ (NGC~300), $\sim$52$\times$ 52 pc$^{2}$ (NGC~7793) and  $\sim$75.5$\times$ 75.5 pc$^{2}$ (NGC~5068).
    The large central circles indicate the objects which ID correspond to the number in the upper left corner, while other objects of the dusty young population in the same field are shown by smaller symbols: circles for point-source dominated sources within SFRs, squares for point-source dominated isolated sources, and pentagons for compact marginally resolved sources.}
    \label{fig:YSO_isolated}
\end{figure*}

\begin{figure*}
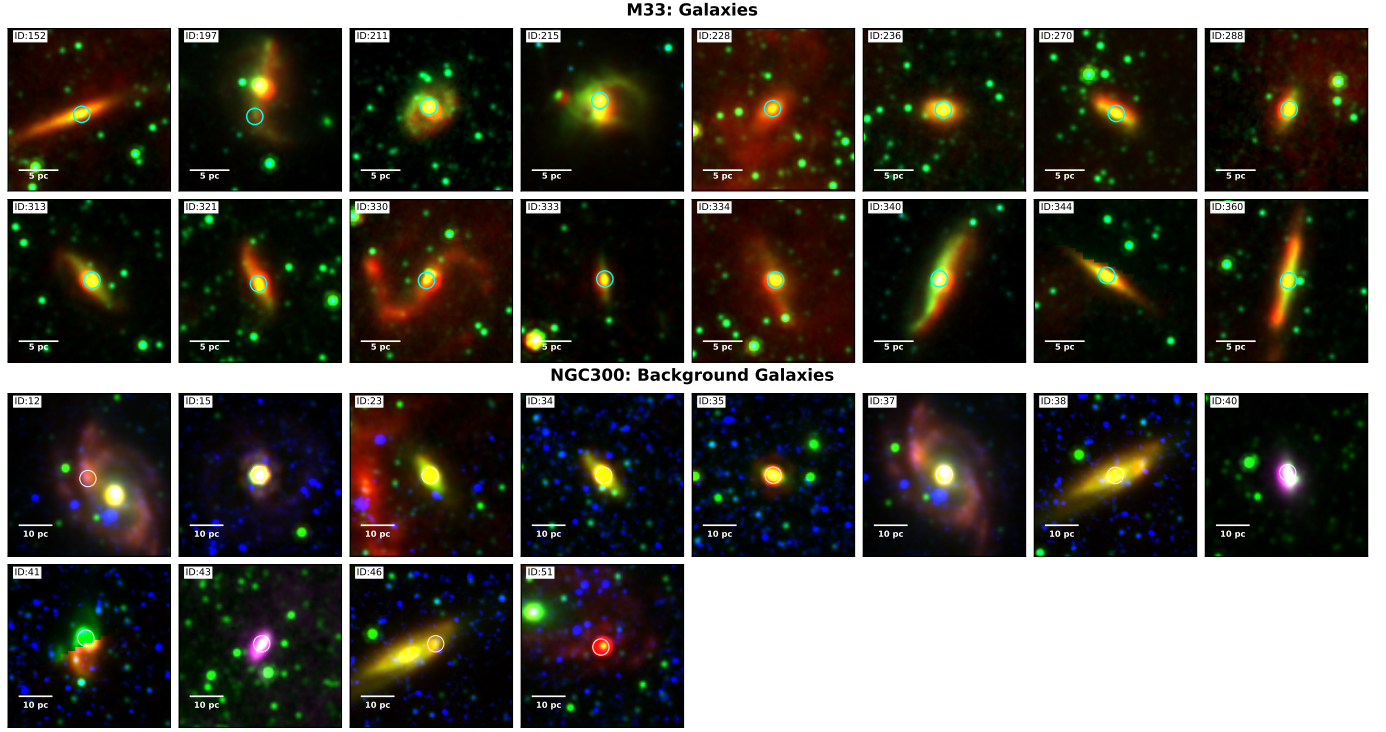

    \centering
    \largefig{\includegraphics[width=1\linewidth]{categories_galaxies_RGB_10_4_k_m33_5arc_examples.pdf}\\
    \includegraphics[width=1\linewidth]{categories_Gal_RGB_ngc0300_5.pdf}}
    \caption{Examples of background galaxies.
    Blue:F475W (M~33)/F435W (NGC~300), Green:F430M, Red: F1000W. Cutout are 5$\times$5 arcsec$^{2}$ which is equivalent to $\sim$20$\times$ 20 pc$^{2}$ at the distance of M~33, $\sim$49$\times$ 49 pc$^{2}$ (NGC~300), and  $\sim$87$\times$ 87 pc$^{2}$ (NGC~7793).}
    \label{fig:GAL_M33}
\end{figure*}

Examples of objects in the categories discussed in Section~\ref{subsec:classification} are presented in Figures~\ref{fig:YSO_SFR}-~\ref{fig:CLTs}: 
\begin{itemize}
\item Point-source dominated within SFRs (Fig.~\ref{fig:YSO_SFR})
\item isolated point-source dominated (Fig.~\ref{fig:YSO_isolated})
\item Compact marginally resolved (Fig.~\ref{fig:CLTs})
\item background galaxies (Fig.~\ref{fig:GAL_M33})
\end{itemize}

 \begin{figure*}
    \centering
    \largefig{\includegraphics[width=1\linewidth]{categories_Marginally_resolved_RGB_10_4_2_m33_examples.pdf}\\
        \includegraphics[width=1\linewidth]{categories_Marginally_resolved_RGB_10_4_2_ngc0300_example.pdf}\\
    \includegraphics[width=1\linewidth]{categories_marginalli_resolved_RGB_10_4_2_ngc7793_examples.pdf}\\
        \includegraphics[width=1\linewidth]{categories_marginally_resolved_RGB_10_4_2_ngc5068_examples.pdf}}
    \caption{Examples of compact marginally resolved sources.
    Blue: F210M (M~33) or F200W (NGC~300,NGC~7793 and NGC~7793) / Green:F430M (M~33), F444W (NGC~300 and NGC~7793) or F360M (NGC~5068)/, Red: F1000W. Cutout are 3$\times$3 arcsec$^{2}$ which is equivalent to $\sim$13.6$\times$ 13.6 pc$^{2}$ at the distance of M~33, $\sim$30$\times$ 30 pc$^{2}$ (NGC~300), $\sim$52$\times$ 52 pc$^{2}$ (NGC~7793) and  $\sim$75.5$\times$ 75.5 pc$^{2}$ (NGC~5068).
    The large central circles indicate the objects which ID correspond to the number in the upper left corner, while other objects of the dusty young population in the same field are shown by smaller symbols: circles for point-source dominated sources within SFRs, squares for point-source dominated isolated sources, and pentagons for compact marginally resolved sources.}
    \label{fig:CLTs}
\end{figure*}


\section{Color correction between JWST and Spitzer filters}
\label{append:color_corr}



To enable comparisons between MYSO studies based on JWST and earlier work using Spitzer, we derive color corrections between filters from the two facilities. The SAGE-Spec survey \citep{sage-spec} provides mid-infrared Spitzer IRS spectroscopy from 5.3 to 38 $\mu$m with spectral resolutions R$\sim$60–600 for a range of Magellanic sources, including evolved dusty stars, massive young stellar objects, planetary nebulae, and supernova remnants. These spectra allow us to compute synthetic photometry in the relevant JWST/MIRI bandpasses and establish relationships between the Spitzer and JWST filter systems.

\begin{figure}
    \centering
    \includegraphics[width=0.3\linewidth]{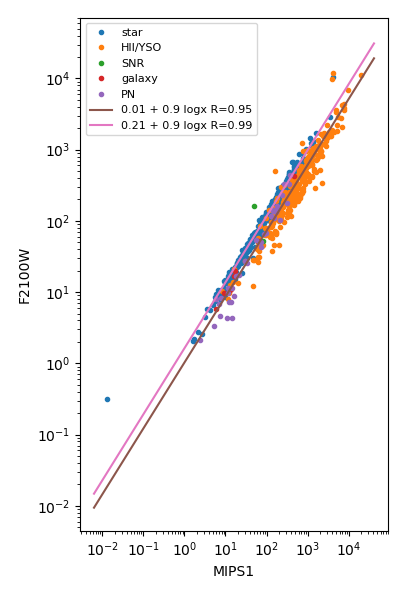}
    \includegraphics[width=0.3\linewidth]{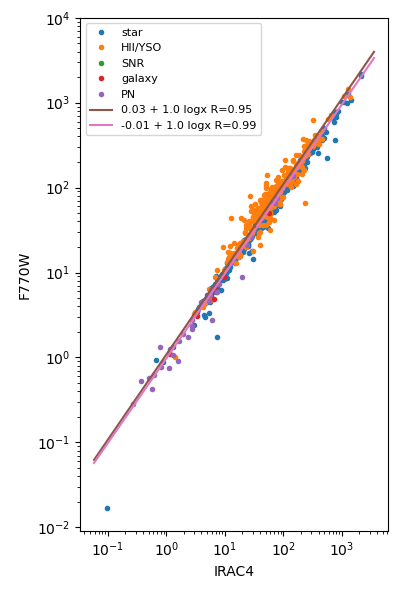}
    \includegraphics[width=0.3\linewidth]{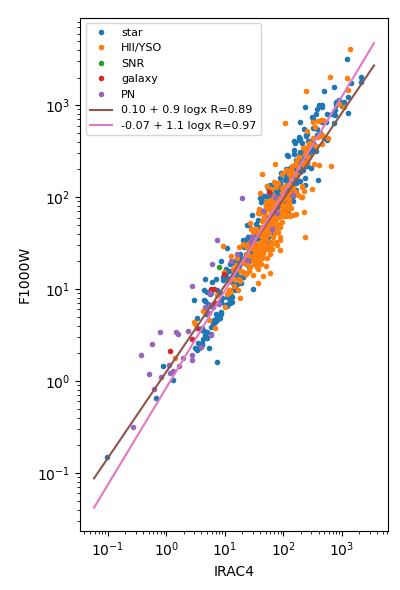}
    \caption{Comparison of synthetic JWST F770W, F1000W, and F2100W fluxes and corresponding Spitzer IRAC4 (7.87 $\mu$m) and MIPS1 (23.7 $\mu$m) photometric fluxes derived from Spitzer IRS spectra for dusty Magellanic sources in the SAGE-Spec survey (evolved stars, MYSOs, and a small number of PNe and SNRs). The panels show log(F$_\mathrm{JWST}$) versus log(F$_\mathrm{Spitzer}$) for each matched filter pair; flux densities are F$_\nu$ in mJy. Linear fits to the log–log relations are overplotted, both for the entire sample shown, and the subset of YSO and dusty HII regions. The fitted slopes, intercepts, and Spearman correlation coefficients for each comparison are listed in the legend. This comparison quantifies the consistency between Spitzer- and JWST-based flux measurements across diverse dusty sources and provides the empirical basis for the color corrections used to relate Spitzer MYSO studies to the JWST results presented here.}
    \label{fig:spitzer_mag_corrections}
\end{figure}

We compute flux densities in the JWST F770W, F1000W, and F2100W MIRI filters as in  \citet{Jones2017}, and compare them with Spitzer IRAC4 (7.87 $\mu$m) and MIPS1 (23.7 $\mu$m) photometry (Fig.~\ref{fig:spitzer_mag_corrections}). Because Spitzer did not have a filter closely matching JWST/F1000W, we compare F1000W flux densities to those measured in IRAC4. The resulting correlations are tight: both F770W:IRAC4 and F2100W:MIPS1 pairs show slopes and Spearman coefficients near unity, indicating very similar flux densities across the two missions for these wavelengths.

\begin{figure}
    \centering
    \includegraphics[width=0.3\linewidth]{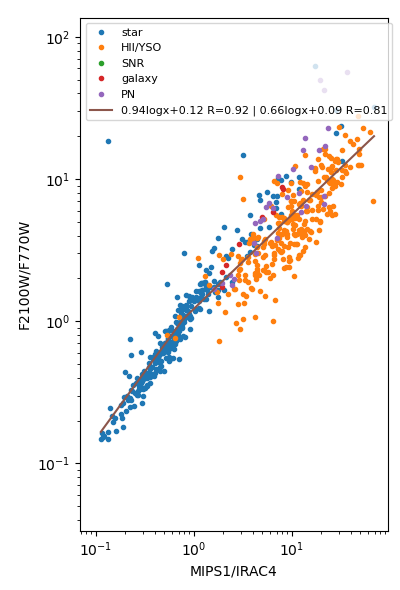}
    \includegraphics[width=0.3\linewidth]{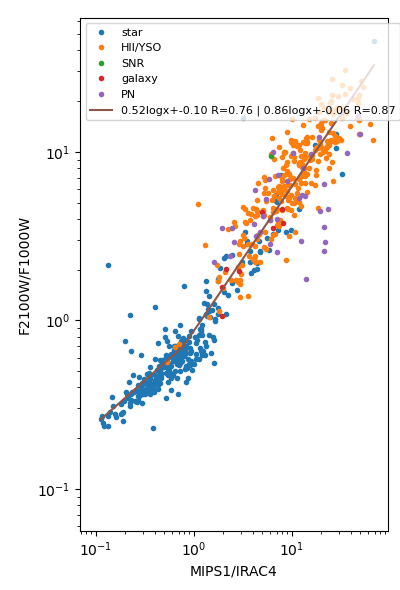}
    \caption{Comparison of colors computed from synthetic JWST fluxes with Spitzer photometry. The two panels show the relationships between F2100W/F1000W and F2100W/F770W  with MIPS1/IRAC4.  Linear fits are overplotted and the corresponding slopes, intercepts, and Spearman correlation coefficients listed in the legend. These relations provide the color transformations used to translate Spitzer-based CMDs to the JWST filter system.}
    \label{fig:spitzer_col_corrections}
\end{figure}

Figure~\ref{fig:spitzer_col_corrections} shows the relationships between the colors. A pure Rayleigh–Jeans spectrum shifts redward by approximately a factor of two when moving from IRAC4/MIPS1 colors to F1000W/F2100W colors, simply because the JWST filters are closer together in wavelength. The bluest sources in the sample follow this expectation, while dust-dominated sources exhibit much smaller color shifts. Consequently, when comparing Spitzer-based color–magnitude diagrams (e.g., Fig.~\ref{fig:lmc_cmd_10_24}) to JWST diagrams, main-sequence stars should be shifted redward by $\sim$2, whereas dusty sources (such as MYSOs) remain at nearly the same colors. These empirically derived corrections allow consistent interpretation of MYSO populations across Spitzer and JWST studies.

\section*{Acknowledgements}

The authors thank the referee for the careful reading of the manuscript and for the constructive comments and suggestions, which helped to improve the clarity and quality of this work.
This work is based on observations made with the NASA/ESA/CSA James Webb Space Telescope (program \#2130 and \#2107) and the NASA/ESA Hubble Space Telescope (program \#15654).   
The data were obtained from the Mikulski Archive for Space Telescopes at the Space Telescope Science Institute, which is operated by the Association of Universities for Research in Astronomy, Inc., under NASA contract NAS 5-03127 for JWST and 5-26555 for HST.

RSK acknowledges financial support from the ERC via Synergy Grant ``ECOGAL'' (project ID 855130) and from the German Excellence Strategy via the Heidelberg Cluster ``STRUCTURES'' (EXC 2181 - 390900948). In addition RSK is grateful for funding from the BMWE in project ``MAINN'' (funding ID 50OO2206), and from DFG and ANR for project ``STARCLUSTERS'' (funding ID KL 1358/22-1).  

\bibliography{biblio}{}
\bibliographystyle{aasjournal}



\end{document}